\documentclass[aps,a4paper, preprint]{revtex4-1}
\usepackage{graphicx}
\usepackage{amsmath}
\usepackage{bbold}
\usepackage{amssymb}
\usepackage{bm}
\usepackage{times}
\usepackage[utf8]{inputenc}
\usepackage[OT1]{fontenc}
\usepackage{helvet}
\usepackage{placeins}
\usepackage{makecell}
\usepackage{dsfont}
\usepackage{upgreek}
\usepackage[version-1-compatibility]{siunitx}
\usepackage[pdfborder=0]{hyperref}
\hypersetup{
    colorlinks=true,
    linkcolor=blue,          
    citecolor=blue           
}
\newcommand{\orcid}[1]{\href{https://orcid.org/#1}{\includegraphics[width=10pt]{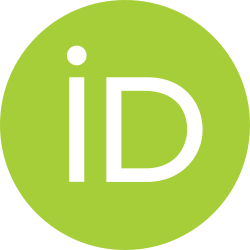}}}

\def\imath{\mathrm{i}}
\def\ee{\mathrm{e}}
\def\beq{\begin{equation}}
\def\eeq{\end{equation}}
\def\beqna{\begin{eqnarray}}
\def\eeqna{\end{eqnarray}}
\def\bea{\begin{array}}
\def\ea{\end{array}}

\def\mg{{\mathcal G}}
\def\MU{{\mathcal U}}
\def\MV{{\mathcal V}}

\def\Re{\operatorname{Re}}
\def\Im{\operatorname{Im}}
\def\etal{{\it et al.~}}

\begin{document}
\title{On deep squeezing or cooling a parametric resonator using feedback}
\author{Adriano A. Batista$^1$\orcid{0000-0003-1781-6455}}
\email{adriano@df.ufcg.edu.br}
\author{Raoni S. N. Moreira$^2$\orcid{0000-0002-9495-1960}, and A. A. Lisboa de Souza$^3$\orcid{0000-0001-5094-6847}}
\affiliation{
$^1$ Departamento de Física, Universidade Federal de Campina Grande\\
Campina Grande-PB, CEP: 58429-900, Brazil\\
$^2$ Departamento de Física, Universidade Federal de Pernambuco, 50670-901 Recife-PE, Brazil\\
$^3$ Departamento de Engenharia Elétrica, Universidade Federal da Paraíba\\
João Pessoa-PB, CEP: 58.051-970, Brazil
}
\date{\today}
\begin{abstract}
Here we analyze ways to achieve deep subthreshold squeezing or
cooling of fluctuations in a parametric resonator enhanced by a lock-in
amplifier feedback loop.
The main advance and novelty of this paper is that we analyzed the full impact
of this feedback on the response of the parametric resonator to an ac
drive or to additive noise.
Due to this feedback, the dynamics of the parametric resonator becomes more
complex and, in addition to a saddle-node bifurcation (SNB), a Hopf bifurcation (HB) at the instability threshold can also occur.
We  calculate the phase-dependent amplification gain in the
response of the resonator to an added ac signal with three independent
techniques: approximately using the averaging method (AM) or the harmonic
balance method (HBM), and more precisely using Floquet theory (FT) and
Green's functions.
The HB was predicted by the AM, FT, and a modified version of the HBM that takes
into account quasi-periodicity.
In our analysis of fluctuations, we were able to calculate the
noise spectral density and the squeezing or cooling of fluctuations using
Fourier techniques and FT.
Deamplification and cooling occur near the HB, whereas deep squeezing and
heating occur near a SNB.
Furthermore, we show that there is an optimal amount of parametric pumping and
feedback to reach the strongest cooling.
The theoretical framework presented here is general and could be applied to a
wide range of physical implementations.
\end{abstract}
\maketitle
\section{Introduction}
Parametric modulation offers a powerful route for suppressing fluctuations in
resonators across physics and engineering \cite{bachtold2022mesoscopic},
enabling both phase-dependent amplification and squeezing. 
It has been used to enhance the sensitivity of gravitational-wave detectors
\cite{Aasi2013ligo, dwyer2022squeezing}, for preparing nonclassical states of
mechanical resonators \cite{aspelmeyer2014cavity, you2017strong, Marti2024}, for
implementing very sensitive accelerometers \cite{zhao2019toward}, force
\cite{moser2013ultrasensitive} and mass sensors \cite{zhang2005application,
papariello2016ultrasensitive}.
More recently, parametric modulation has played a fundamental role in the
implementation of phase state qubits in Kerr parametric oscillators
\cite{grimm2020stabilization, shen2026readout}, in qubit readout in circuit
quantum electrodynamics \cite{xu2025parametric,xu2026onchip}, and 
in quantum sensing \cite{kamba2025quantumSqueezing}.
Despite of these advances, conventional parametric schemes face two fundamental
limits: squeezing is bounded to $-6~\si{dB}$ at the instability threshold
\cite{rugar91,cleland2005thermomechanical,batista2025deep} and cooling is not possible.
Feedback control provides an alternate strategy without these limitations, but
its combination with parametric modulation has remained largely unexplored from
a theoretical standpoint. 
In this work, we fill that gap by analyzing a single-degree-of-freedom (SDOF)
parametric resonator augmented with a lock-in amplifier (LIA) feedback loop that
can be tuned for either squeezing or cooling. 

The thermo-mechanical squeezing limitation was overcome only in 2013 by Vinante
and Falferi \cite{vinante2013feedback}. 
In their work, a $-11.3~\si{dB}$ in squeezing of thermal fluctuations of a
parametrically modulated microscopic silicon beam cantilever was reached
experimentally with the assistance of feedback.
Further experimental advances using the same squeezing technique with
parametric pump and feedback were subsequently made
by Poot \etal \cite{poot2015deep}, Sonar \etal \cite{sonar2018strong},
and Mashaal \etal \cite{mashaal2025strong}.
Although these papers advanced the experimental implementation of
thermomechanical squeezing
enhanced by feedback, they did not advance the theoretical analysis of
fluctuations developed by Vinante and Falferi. Their model oversimplifies
the dynamics of the parametric resonator with feedback by averaging the
equations of motion from the start, what reduced the dynamical system dimensions
to 2.
Consequently, this led to their missing of a Hopf bifurcation (HB).
In addition to that, they did not provide a way to calculate the noise spectral density (NSD)
of the parametric resonator fluctuations nor the NSD of the LIA outputs.

Another method used to reduce fluctuations is cooling.
It can be implemented in (opto)mechanical resonators in several ways: parametric
feedback 
\cite{Briant2003optical, Gieseler2012subkelvin, Gao2024feedback}, feedback 
\cite{Hopkins2003feedback, poggio2007feedback, Guo2019feedback,deli2020Cooling}, sideband
 \cite{Schliesser2009resolved}, passive cooling using nuclear demagnetization
\cite{vanEverdingen2026}, and radiation pressure 
\cite{Arcizet2006radiation}.
Effective temperatures as low as a few millikelvin can be achieved using these
feedback schemes.
Even quantum ground state cooling has been implemented experimentally in
nanomechanical resonators \cite{OConnel2010quantum}.
The comprehensive review of cavity optomechanics by
Aspelmeyer \etal \cite{aspelmeyer2014cavity}
has many more citations on cooling resonators.
The feedback scheme analyzed here is capable of achieving cooling and it could
be applied and tested in various types of physical systems.
 
Here we extend our previous works on fluctuations and on squeezing of
fluctuations in parametrically-modulated resonators with added noise
\cite{batista2011signal, batista2024amplification, batista2025deep} by including feedback.
Specifically, we analyze a linear feedback scheme inspired by the one proposed
by Vinante and Falferi to cool or enhance squeezing of fluctuations.
When the parametric resonator with feedback is excited by an added ac signal,
without noise, we analytically obtain the gain curve as a function of phase
using the averaging method (AM) and the harmonic balance method (HBM).
We show that the SDOF parametric resonator with
feedback has a higher dynamical dimensionality, with three Floquet multipliers
(FMs),
what allows HBs.
We find that this is a new route to instability, in addition to period-doubling
or saddle-node bifurcations (SNBs). 
The analytical results of this approach are verified with numerical results
based on Floquet theory (FT).
This analysis of a deterministic stationary response complements our stochastic
analysis of squeezing and cooling.
It is a way to independently check our results.
Subsequently, in our analysis of fluctuations, we avoid directly averaging the
stochastic differential equations. 
Instead, we use the Green's functions method in the frequency domain.
We obtain the Green's functions, approximately, using a perturbative Fourier
transform method or, more exactly, using FT.
With this method, we calculate the NSD of the resonator
response to the added noise.
We also obtain the dispersions in the sine and cosine quadratures at half the
pump frequency and the correlation that arises between these quadratures.
Our results go beyond the $-6~\si{dB}$ limit, predict a new
instability route via a HB, and provide a consistent stochastic
theory that has been missing in previous experimental work on feedback‑enhanced
squeezing.

The remainder of this paper is organized as follows.
In Sec.~\ref{Lockin_feedback} we initially analyze degenerate parametric
amplification of an ac signal enhanced by the proposed LIA
 feedback scheme using the AM, the HBM, and FT. 
We also predict the gain, the SNB and the HB lines analytically.
In Sec.~\ref{Noise} we analyze the corresponding stochastic model in which the
ac signal is replaced by white noise.
We calculate the NSD and obtain analytical expressions for the enhanced
squeezing or cooling of fluctuations.
In Sec.~\ref{results} we analyze and discuss our numerical results, and in
Sec.~\ref{conclusion} we draw our conclusions.
In the appendix, we describe the operation of a LIA amplifier, develop
analytic expressions for the gain and for the bifurcation lines based on the HBM.
\section{Parametric amplification with lock-in feedback}
\label{Lockin_feedback}
In the first part of our investigation, we analyze degenerate
parametric amplification enhanced by LIA feedback in a SDOF parametric
amplifier.
The objective being to obtain threshold lines to instability and the gain curve as a function of the signal phase.
This analysis of a deterministic stationary response gives us intuition on our
subsequent analysis of stochastic processes such as fluctuations squeezing and
cooling.
It is a simple way to verify if deep squeezing and cooling are possible.

The equations of motion of the parametric amplifier enhanced by the feedback
scheme are given by the following integro-differential equation
\beq
\ddot x+\gamma \dot x+\omega_0^2x-F_\mathrm{p}\sin(2\omega t)x=
2\eta \sin(\omega t)x_\mathrm{L}(t)
+F_\mathrm{s}\cos(\omega_\mathrm{s} t+\varphi_0),
\label{pa_lockin_feedback_amp}
\eeq
where $\gamma$ is the dissipation coefficient, $\omega_0$ is the angular
natural frequency of the resonator, $F_\mathrm{p}$ is the pump amplitude, $2\omega$ is
the pump frequency, $\eta$ is the feedback constant, $F_\mathrm{s}$ is the input signal
amplitude, $\omega_\mathrm{s}$ is the input signal angular frequency,  $\varphi_0$ is
the input signal phase, and 
\beq
x_\mathrm{L}(t)=\frac1\tau \int_{-\infty}^t \ee^{-(t-t')/\tau}\cos(\omega t')x(t')\,dt'
\label{eq:x_L}
\eeq
is the first-order $RC$ filter output from the cosine quadrature of a LIA
(see the appendix for more details).
Here, $\tau$ is the LIA time constant.
The above integro-differential equation can be rewritten as
the following non-autonomous ordinary differential equation (ODE) system
\beq
\begin{aligned}
\dot x&=y,\\
\dot y&=-\omega_0^2x-\gamma y+F_\mathrm{p}\sin(2\omega t)x
+ 2\eta\sin(\omega t)z+F_\mathrm{s}\cos(\omega_\mathrm{s} t+\varphi_0),\\
\dot z&=-\frac z\tau+\cos(\omega t)\frac x\tau,\\
\end{aligned}
\label{pa_lockin_feedback_amp2}
\eeq
where $z(t)=x_\mathrm{L}(t)$.

\FloatBarrier
\subsection{The AM analysis of degenerate parametric amplification}
\label{sec:AM}
Let us analyze the amplification dependence on phase obtained from
Eqs.~\eqref{pa_lockin_feedback_amp2} in a similar way to what was performed by
Rugar and Grütter \cite{rugar91}.
This analysis provides us with insight on the response of this system to added
noise.  We now transform the fast variables $(x(t), y(t))$ from
Eq.~\eqref{pa_lockin_feedback_amp2} into the slow variables $(\MU(t), \MV(t))$
via $\MU(t)= x(t)\cos(\omega t)-\frac{y(t)}\omega\sin(\omega t)$ and $\MV(t)=
-x(t)\sin(\omega t)-\frac{y(t)}\omega\cos(\omega t)$.
As long as $\omega_0\tau>>1$, the variable $z$ varies slowly, hence it does
not need to be transformed.
We find the equations of motion for $\MU(t)$ and $\MV(t)$ to be
\beq
\left(\bea{c}
\dot \MU \\ \dot \MV
\ea
\right)
=
-\frac{1}{\omega}\left(\bea{c} \sin(\omega t) f(x,y,z,t)\\
\cos(\omega t) f(x,y,z, t)\ea\right),
\eeq
where $f(x,y,z, t)= -\Omega x+F_\mathrm{p}\sin(2\omega t)\;x-\gamma y+2\eta
z\sin(\omega t)+F_\mathrm{s}\cos(\omega t+\varphi(t))$, with $\Omega=\omega_0^2-\omega^2$ and $\varphi(t)=\delta t+\varphi_0$, with $\delta=\omega_\mathrm{s}-\omega$.
Subsequently, we apply the AM to these equations and obtain the
following averaged dynamical system
\beq
\begin{aligned}
\dot u &=
-\frac1{2\omega}\left[\left(\frac{F_\mathrm{p}}2+\gamma\omega\right)u+\Omega v+2\eta z
-F_\mathrm{s}\sin\varphi(t)\right],\\
\dot v &= \frac1{2\omega}\left[\Omega
u+\left(\frac{F_\mathrm{p}}2-\gamma\omega\right)v-F_\mathrm{s}\cos\varphi(t)\right],\\
\dot z&=\frac{u}{2\tau}-\frac z\tau,
\end{aligned}
\label{uvz_avg}
\eeq
where $\MU(t)\approx u(t)$ and $\MV(t)\approx v(t)$.
When $\omega=\omega_\mathrm{s}$, we can write the fixed-point solution of
Eq.~\eqref{uvz_avg} as
\beq
\left(
\bea{c}
u\\v
\ea
\right)
=\frac{F_\mathrm{s}}{\left(\frac{F_\mathrm{p}}2+\gamma\omega
+\eta\right)\left(\frac{F_\mathrm{p}}2-\gamma\omega\right)-\Omega^2}
\left(
\bea{cc}
\frac{F_\mathrm{p}}2-\gamma\omega&-\Omega \\
 -\Omega &\frac{F_\mathrm{p}}2+\gamma\omega +\eta
 \ea
\right)
\left(
\bea{c}
\sin\varphi_0\\
\cos\varphi_0
\ea
\right).
\eeq
Consequently, we can write the gain ($r/|\chi(\omega)F_\mathrm{s}|$) in the averaging approximation as
\beq
\resizebox{.92\hsize}{!}{$
{\small G_{\mathrm{avg}}(\omega,\varphi_0)}=
\frac{\sqrt{\Omega^2+\frac{F_\mathrm{p}^2}4+\gamma^2\omega^2+\frac{\eta^2}2+\eta\left(\frac{F_\mathrm{p}}2+\gamma\omega\right)+\left[F_\mathrm{p}\gamma\omega+\frac\eta2\left(\eta+F_\mathrm{p}+2\gamma\omega\right)\right]\cos(2\varphi_0)-\Omega(F_\mathrm{p}+\eta)\sin(2\varphi_0)}}
{\left|\left(\frac{F_\mathrm{p}}2+\gamma\omega
+\eta\right)\left(\frac{F_\mathrm{p}}2-\gamma\omega\right)-\Omega^2\right||\chi(\omega)|},
\label{deep_squeeze}
$}
\eeq
where  $r=\sqrt{u^2+v^2}$ and
$\chi(\omega)=\frac1{\Omega-\imath\gamma\omega}$ is the susceptibility.
From Eq.~\eqref{deep_squeeze}, we find the minimum and maximum gain of
the response of the parametric amplifier at zero detuning ($\Omega=0$) to be
\beq
\begin{aligned}
G_{\mathrm{avg}}^{\mathrm{min}}&=\frac{r_{\mathrm{min}}}{r_0} = \frac{\gamma\omega}{\left|F_\mathrm{p}/2+\gamma\omega+\eta\right|},\\
G_{\mathrm{avg}}^{\mathrm{max}}&=\frac{r_{\mathrm{max}}}{r_0} = \frac{\gamma\omega}{\left|F_\mathrm{p}/2-\gamma\omega\right|},
\end{aligned}
\label{r_min_r_max}
\eeq
where $r_0=|\chi(\omega)F_\mathrm{s}|$ is the forced harmonic oscillator amplitude.
\subsubsection{Stability analysis}
At an instability line, one has to investigate the eigenvalues of the Jacobian
matrix of the dynamical system given in Eq. (5) to decide on what kind of
bifurcation to instability one has.
The Jacobian matrix is given by
\beq
M =
\begin{pmatrix}
-\frac1{2\omega}\left(\frac{F_\mathrm{p}}2+\gamma\omega\right)&-\frac\Omega{2\omega}
&-\frac\eta\omega\\
 \frac\Omega{2\omega} &\frac1{2\omega}\left(\frac{F_\mathrm{p}}2-\gamma\omega\right)&0\\
\frac1{2\tau}&0&-\frac1\tau
\end{pmatrix}.
\eeq
We find the eigenvalues by solving the following characteristic equation
$\det (M-\lambda\mathbb{1})=0$, that is
\beq
\left(\lambda+\frac1\tau\right)\left[\left(\frac{F_p}{4\omega}\right)^2
-\left(\lambda+\frac\gamma2\right)^2-\frac{\Omega^2}{4\omega^2}\right]
-\frac\eta{2\omega\tau}\left[\lambda-\frac1{2\omega}\left(\frac{F_p}2-\gamma\omega\right)\right]=0.
\label{eq:character0}
\eeq
We can rearrange this equation in the following format
\beq
\lambda^3+\left(\frac1\tau+\gamma\right)\lambda^2+\left[\frac\gamma\tau+\frac1{4\omega^2}\left(\gamma^2\omega^2-\frac{F_\mathrm{p}^2}4\right)+\frac{\Omega^2}{4\omega^2}+\frac\eta{2\omega\tau}\right]\lambda-\det M=0, 
\label{eq:character}
\eeq
where
\[
\det
M=\frac1{4\omega^2\tau}\left[\left(\frac{F_\mathrm{p}}2+\gamma\omega+\eta\right)\left(\frac{F_\mathrm{p}}2-\gamma\omega\right)-\Omega^2\right].
\]
When $\det M=0$, we necessarily have a $\lambda=0$ eigenvalue.
In this case, the bifurcation is a SNB.
The SNB thresholds for parametric oscillations with lock-in feedback are given
by
\beq
F_\mathrm{p}^2+2\eta F_\mathrm{p}-4\left(\gamma^2\omega^2+\Omega^2+\eta\gamma\omega\right)=0
\implies
F_\mathrm{p}=-\eta\pm\sqrt{4\Omega^2+(\eta+2\gamma\omega)^2}.
\label{eq:threshold_avg}
\eeq

On the other hand, at a HB, $\det M\neq0$, we
must have a pair of purely imaginary eigenvalues $\lambda_\pm=\pm i\Delta$, with
$\Delta\in\mathbb{R}$.
Plugging this into the characteristic equation \eqref{eq:character}, we obtain
the necessary condition for a HB.
In our system, this is the following quadratic equation in $F_p$ 
\beq
\Delta^2=\frac{-\det M}{\left(\frac1\tau+\gamma\right)}
=\left[\frac\gamma\tau+\frac1{4\omega^2}\left(\gamma^2\omega^2-\frac{F_\mathrm{p}^2}4\right)+\frac{\Omega^2}{4\omega^2}+\frac\eta{2\omega\tau}\right].
\label{eq:Delta_avg}
\eeq
Explicitly, for $\eta\neq0$, this equation can be written as
\beq
\frac{\gamma}{16\omega^2}F_\mathrm{p}^2-\frac{\eta}{8\omega^2\tau}F_\mathrm{p}+\frac{\eta\gamma\omega+\Omega^2}{4\omega^2\tau}-\frac\gamma{\tau}\left(\frac1\tau+\gamma\right)-\frac{\gamma^3}4=0.
\eeq
This predicts the following two HB lines
\beq
F_\mathrm{p} = 
\frac\eta{\gamma\tau} \pm 
\sqrt{
\left(\frac{2\gamma^2\omega\tau + 4\gamma\omega + \eta}{\gamma\tau}\right)^2
+ 4\Omega^2}.
\label{eq:Hopf_line_avg}
\eeq
The positive branch  is deep inside the
instability region but the other branch agrees with the modified HBM and the Floquet HB line.
If however, $\eta=0$, a HB is not possible.
This can be seen in Eq.~\eqref{eq:character0}, where there is no imaginary
root, or in Eq.~\eqref{eq:Delta_avg}, where we find that $\Delta^2<0$, a
contradiction with the assumption that $\Delta$ is real.
\subsection{Floquet theory analysis}
Before we apply FT to analyze parametric amplification, we review the basic
features of FT developed in Ref.~\cite{batista2024amplification}.
Given the dynamical system $\dot X={\boldsymbol A}(t)X$,
where $X\in\mathds{R}^n$ and ${\boldsymbol A}(t)\in\mathds{R}^{n\times n}$.
The solution to this system is $X(t)={\boldsymbol\Phi}(t,t_0)X_0$, with 
$X(t_0)=X_0$ and $\boldsymbol\Phi(t_0, t_0)=\mathbb{1}$.
The $n\times n$ time-evolution operator $\boldsymbol\Phi(t, t_0)$ obeys
$\dot{\boldsymbol\Phi}(t, t_0)=\boldsymbol A(t)\boldsymbol\Phi(t,t_0)$.
By Floquet's theorem, when $\boldsymbol A(t+T)=\boldsymbol A(t)\implies$ $\boldsymbol\Phi(t,
t_0)={\boldsymbol P}(t-t_0)\ee^{\boldsymbol B(t-t_0)}$, where $\boldsymbol P(t+T)=\boldsymbol P(t)$. 
Hence, we find $\boldsymbol\Phi(t'+T, t')=\boldsymbol\Phi(t_0+T,
t_0)=\ee^{{\boldsymbol B}T}$.
The eigenvalues of $\boldsymbol\Phi(T,0)$ are known as FMs ($\mu_i$)
and the eigenvalues of ${\boldsymbol B}$ are known as Floquet exponents (FEs).

The stationary solution of the inhomogeneous real $n$-dimensional ODE system
\beq
\dot X=\boldsymbol A(t)X+F(t)
\eeq 
is given by
\beq
X(t)=\int_{t_0}^t\boldsymbol G(t, t')F(t')dt',
\label{x_t}
\eeq
where $\boldsymbol G(t, t')=\boldsymbol\Phi(t, t_0)\boldsymbol\Phi^{-1}(t',
t_0)=\boldsymbol P(t-t_0)\ee^{\boldsymbol B(t-t')}\boldsymbol Q(t'-t_0)$ is the
generalized Green's function, $\boldsymbol Q(t')=\boldsymbol P^{-1}(t')$,
and $F(t)$ is the additive excitation vector that in our model could be an ac signal or white noise.
We can write the Fourier series of these periodic matrices as
\beq
\begin{aligned}
\boldsymbol P(t)=\sum_{n=-\infty}^\infty \boldsymbol p_{n}\ee^{\imath n\omega t},\\
\boldsymbol Q(t)=\sum_{n=-\infty}^\infty \boldsymbol q_{n}\ee^{\imath n\omega t},
\end{aligned}
\label{eq:PQ}
\eeq
where we assume that the period is $T=2\uppi/\omega$. 

According to Ref.~\cite{batista2024amplification}, the response in the frequency
domain of the parametrically driven resonator to an added signal is given by
\beq
\tilde X(\nu)\approx{{\tilde{\bm  G_0}}}(\nu)\tilde F(\nu)+{\boldsymbol G_+}(\nu)\tilde F(\nu-2\omega)
+{\boldsymbol G_+^*}(-\nu)\tilde F(\nu+2\omega),
\label{fundEqF}
\eeq
where the tilde symbol over any given function $f(t)$ denotes the Fourier
transform as in $\tilde f(\nu)=\int_{-\infty}^\infty \ee^{\imath\nu t}f(t)dt$.
The matrix Green's functions in the frequency domain
$\tilde{\bm G_0}(\nu)$ and
${\boldsymbol G_+}(\nu)$ are given by
\beq
\begin{aligned}
\tilde{\bm G}_0(\nu)&\approx
-\boldsymbol p_1^*\left[\boldsymbol{B}+\imath(\nu-\omega)\mathbb{1}\right]^{-1}\boldsymbol
q_1-\boldsymbol p_1\left[\boldsymbol{B}+\imath(\nu+\omega)\mathbb{1}\right]^{-1}\boldsymbol q_1^*,\\
{\boldsymbol G_+}(\nu) &= -\boldsymbol p_1^*\left[\boldsymbol B+\imath(\nu-\omega)\mathbb{1}\right]^{-1}\boldsymbol q_1^*.\\
\end{aligned}
\label{G_0G_+}
\eeq
We point out that the dynamical system is stable as long as all FEs have
negative real part and the threshold to instability occurs when there is at
least one eigenvalue with zero real part.
Details on obtaining these functions numerically can be found in Ref.~\cite{batista2024amplification}.
The $i$-th component response is given by
\beq
\tilde X_i(\nu)\approx\sum_j{\tilde G_{0,ij}}(\nu)\tilde F_j(\nu)
+\sum_j{G_{+, ij}}(\nu)\tilde F_j(\nu-2\omega)
+\sum_j{G^*_{+,ij}}(-\nu)\tilde F_j(\nu+2\omega).
\label{fundEqX_iF_j}
\eeq
In the model given in Eq.~\eqref{pa_lockin_feedback_amp2}, we have
\beq
X(t)=
\left(
\bea{c}
X_0(t)\\
X_1(t)\\
X_2(t)
\ea
\right)
=
\left(
\bea{c}
x(t)\\
y(t)\\
z(t)
\ea
\right).
\eeq
\subsubsection{Parametric amplification}
As the amplifier described here is linear in relation to its response to an
added excitation (ac signal or noise), it amplifies both signal and noise in similar ways.
The difference is that the ac signal is coherent and is a simple tone,
while the noise is incoherent and has a very broad constant spectrum.
The strongest responses of the amplifier occur at degenerate parametric
amplification.
The strongest response to noise also occurs at the spectral components near
$\omega$, at half the parametric pump frequency due to the parametric
conversions (up and down).
Hence, analyzing the phase-dependent degenerate parametric amplification will
give us insight on what type of response to noise we will obtain: deep squeezing, squeezing, cooling, or heating. 
We point out that this approach of analyzing phase-dependent degenerate
parametric amplification already appeared in the paper by Rugar and Grütter
\cite{rugar91}.

Here the additive excitation vector is given by 
\beq
F(t)=
\left(
\bea{c}
0\\
\cos(\omega_\mathrm{s}t+\varphi_0)\\
0
\ea
\right)
\implies
\tilde F(\nu)=
\uppi\left(
\bea{c}
0\\
\ee^{\imath\varphi_0}\delta(\nu+\omega_\mathrm{s})+\ee^{-\imath\varphi_0}\delta(\nu-\omega_\mathrm{s})\\
0
\ea
\right).
\eeq
The response of the parametric amplifier is given by
\beq
\begin{aligned}
\tilde x(\nu)&=\tilde G_{0,01}(\nu)\tilde F_1(\nu)
+G_{+, 01}(\nu)\tilde F_1(\nu-2\omega)
+G^*_{+,01}(-\nu)\tilde F_1(\nu+2\omega)\\
&=\uppi\left[\tilde G_{0,01}(\omega_\mathrm{s})\ee^{-\imath\varphi_0}\delta(\nu-\omega_\mathrm{s})
+\tilde G_{0,01}(-\omega_\mathrm{s})\ee^{\imath\varphi_0}\delta(\nu+\omega_\mathrm{s}) \right]\\
&+\uppi\left[G_{+, 01}(2\omega+\omega_\mathrm{s})\ee^{-\imath\varphi_0}\delta(\nu-2\omega-\omega_\mathrm{s})
+G_{+, 01}(2\omega-\omega_\mathrm{s})e^{\imath\varphi_0}\delta(\nu-2\omega+\omega_\mathrm{s})\right]\\
&+\uppi\left[G^*_{+,01}(2\omega+\omega_\mathrm{s})\ee^{\imath\varphi_0}\delta(\nu+2\omega+\omega_\mathrm{s})
+G^*_{+,01}(2\omega-\omega_\mathrm{s})\ee^{-\imath\varphi_0}\delta(\nu+2\omega-\omega_\mathrm{s})\right].
\end{aligned}
\eeq
We then Fourier transform the above equation back to the time domain and find
\beq
\resizebox{.91\hsize}{!}{$
\begin{aligned}
x(t)&=\frac12\left[\tilde G_{0,01}(\omega_\mathrm{s})\ee^{-\imath\varphi_0}
e^{-\imath\omega_\mathrm{s}t}
+\tilde G_{0,01}(-\omega_\mathrm{s})e^{\imath\varphi_0}\ee^{\imath\omega_\mathrm{s}t} \right]\\
&+\frac12\left[G_{+,
01}(2\omega+\omega_\mathrm{s})\ee^{-\imath\varphi_0}\ee^{-\imath(2\omega+\omega_\mathrm{s})t}
+G_{+, 01}(2\omega-\omega_\mathrm{s})\ee^{\imath\varphi_0}\ee^{-\imath(2\omega-\omega_\mathrm{s})t}\right]\\
&+\frac12\left[G^*_{+,01}(2\omega+\omega_\mathrm{s})\ee^{\imath\varphi_0}\ee^{\imath(2\omega+\omega_\mathrm{s})t}
+G^*_{+,01}(2\omega-\omega_\mathrm{s})\ee^{-\imath\varphi_0}\ee^{\imath(2\omega-\omega_\mathrm{s})t}\right]\\
&\approx\frac12\left[\tilde
G_{0,01}(\omega_\mathrm{s})\ee^{-\imath\varphi_0}\ee^{-\imath\omega_\mathrm{s}t}
+\tilde G_{0,01}(-\omega_\mathrm{s})\ee^{\imath\varphi_0}\ee^{\imath\omega_\mathrm{s}t} \right]
+\frac12\left[
G_{+, 01}(\omega_\mathrm{i})\ee^{\imath\varphi_0}\ee^{-\imath\omega_\mathrm{i}t}
+G^*_{+,01}(\omega_\mathrm{i})\ee^{-\imath\varphi_0}\ee^{\imath\omega_\mathrm{i}t}\right]\\
&=\frac12\left[\tilde G_{0,01}(\omega_\mathrm{s})\ee^{-\imath\varphi}\ee^{-\imath\delta t}
+G_{+, 01}(\omega_\mathrm{i})\ee^{\imath\varphi_0}\ee^{\imath\delta t}
\right]\ee^{-\imath\omega t}
+\frac12\left[
\tilde G^*_{0,01}(\omega_\mathrm{s})\ee^{\imath\varphi_0}\ee^{\imath\delta t}
+G^*_{+,01}(\omega_\mathrm{i})\ee^{-\imath\varphi_0}\ee^{-\imath\delta
t}\right]\ee^{\imath\omega t},
\end{aligned}
$}
\eeq
where we used $\omega_\mathrm{s}=\omega+\delta$ and $\omega_\mathrm{i}=\omega-\delta$. 
Hence, we find that the positive and negative envelopes of $x(t)$ are given by
\beq
\pm\left|\tilde G_{0,01}(\omega_\mathrm{s})
+G_{+, 01}(\omega_\mathrm{i})\ee^{2\imath\varphi_0}\ee^{2\imath\delta t} \right|.
\label{eq:envelopes}
\eeq
At $\delta=0$, we obtain the FT expression for the gain
\beq
G_{\mathrm{FT}}(\varphi_0)=\left|\tilde G_{0,01}(\omega)+G_{+,
01}(\omega)\ee^{2\imath\varphi_0} \right|.
\label{gain_FT_fb}
\eeq
\section{Analysis of fluctuations}
\label{Noise}
Here we present our theoretical model of fluctuations squeezing by applying
LIA feedback to SDOF parametrically-modulated resonators with additive white
noise.
Specifically, we analyze and generalize the linear feedback scheme proposed by
Vinante and Falferi \cite{vinante2013feedback} to enhance squeezing of
fluctuations beyond the $-6~\si{dB}$ maximum limit previously set by Rugar and
Grütter.
Afterwards, we analyze the fluctuations in the frequency domain by Fourier
transforming the time-dependent stochastic differential (or Langevin) equations
of our model into an algebraic stochastic system in the frequency domain.
With this procedure we obtain perturbative approximations to the Green's
functions in the frequency domain.
After this point, we apply the same steps developed in
Ref.~\cite{batista2025deep} to obtain the NSD in addition to the dispersions in
the sine and cosine quadratures plus their correlation at half the pump
frequency.
Finally, we apply FT to obtain more exact Green's functions of
this model and repeat the calculations to obtain the NSD and the squeezing.
\subsection{Fourier transform approach}
\label{Fourier}
The Langevin equation of a damped parametrically driven oscillator with added
noise and with feedback from the cosine output of a LIA  is given by
\beq
\begin{aligned}
\dot x &= y,\\
\dot y&=-\omega_0^2x-\gamma y-F_\mathrm{p}\sin(2\omega t)x
+\frac{2\eta}\tau\sin(\omega t)z+r(t),\\
\dot z&=-\frac z\tau+\cos(\omega t)\frac x\tau,
\end{aligned}
\label{po_lockin_feedback_noise}
\eeq
where $r(t)$ is a zero-mean Gaussian white noise with autocorrelation
$\langle r(t)r(t')\rangle= 2 D\delta(t-t')$, where $D$ is the noise level.
Here we assumed that the LIA has only one-stage low-pass $RC$
filter before its cosine channel output.
After Fourier transforming these equations, we find
\beq
\begin{aligned}
\tilde x(\nu)+\frac{\imath F_\mathrm{p}\chi(\nu)}2\left[\tilde x(\nu+2\omega)-\tilde
x(\nu-2\omega)\right]&=-\frac{\imath\eta\chi(\nu)}\tau[\tilde z(\nu+\omega)-\tilde
z(\nu-\omega)]+\chi(\nu)\tilde r(\nu),\\
\tilde z(\nu) &=\frac{\tilde x(\nu+\omega)+\tilde
x(\nu-\omega)}{2(1-\imath\nu\tau)}.
\end{aligned}
\label{eq:fundamental_noise}
\eeq
By shifting $\nu$ in $\tilde z(\nu)$ by $\pm\omega$ we find
\[
\begin{aligned}
\tilde z(\nu-\omega) &=\frac{\tilde x(\nu)+\tilde
x(\nu-2\omega)}{2(1-\imath(\nu-\omega)\tau)},\\
\tilde z(\nu+\omega) &=\frac{\tilde x(\nu)+\tilde
x(\nu+2\omega)}{2(1-\imath(\nu+\omega)\tau)}.\\
\end{aligned}
\]
Hence, we can now rewrite Eq.~\eqref{eq:fundamental_noise} in the following format
\beq
\alpha(\nu)\tilde x(\nu)+\beta^*(-\nu)\tilde x(\nu-2\omega)+\beta(\nu)\tilde x(\nu+2\omega)=\chi(\nu)\tilde r(\nu),
\label{eq:fundamental_noise2}
\eeq
where
\beq
\begin{aligned}
\alpha(\nu) &=
1+\frac{\imath\eta\chi(\nu)}2\left[\frac1{1-\imath(\nu+\omega)\tau}-\frac{1}{1-\imath(\nu-\omega)\tau}\right],\\
\beta(\nu) &= \frac{\imath\chi(\nu)}2\left[F_\mathrm{p}+\frac\eta{1-\imath(\nu+\omega)\tau}\right].
\end{aligned}
\eeq
After shifting $\nu$ by $\pm2\omega$ in the above equation, we neglected terms
such as $\tilde x(\nu\pm4\omega)$.
This approximation is justified because of the resonant
nature of our physical system, we assume the resonator has a high quality factor
and that $\nu$ is near $\omega$.
In this way, we could replace the terms  $\tilde x(\nu\pm2\omega)$ by
expressions dependent on
$\tilde x(\nu)$ and the noise.
Hence, we can approximate the solution to Eq.~\eqref{eq:fundamental_noise2} by
\beq
\tilde x(\nu)=\tilde\mg_0(\nu)\tilde r(\nu)+\mg_+(\nu)\tilde r(\nu-2\omega)+\mg_-(\nu)\tilde r(\nu+2\omega),
\label{x_nu_fb}
\eeq
where
\beq
\begin{aligned}
\tilde\mg_0(\nu) &= \dfrac{\chi(\nu)}{\alpha(\nu)-\dfrac{\beta^*(-\nu)\beta(\nu-2\omega)}{\alpha(\nu-2\omega)}
-\dfrac{\beta(\nu)\beta^*(-\nu-2\omega)}{\alpha(\nu+2\omega)}},\\
\mg_+(\nu) &=
-\dfrac{\beta^*(-\nu)\chi(\nu-2\omega)}{\alpha(\nu-2\omega)\left[\alpha(\nu)-\dfrac{\beta^*(-\nu)\beta(\nu-2\omega)}{\alpha(\nu-2\omega)}
-\dfrac{\beta(\nu)\beta^*(-\nu-2\omega)}{\alpha(\nu+2\omega)}\right]}, \\
\mg_-(\nu) &=
-\dfrac{\beta(\nu)\chi(\nu+2\omega)}{\alpha(\nu+2\omega)\left[\alpha(\nu)-\dfrac{\beta^*(-\nu)\beta(\nu-2\omega)}{\alpha(\nu-2\omega)}
-\dfrac{\beta(\nu)\beta^*(-\nu-2\omega)}{\alpha(\nu+2\omega)}\right]}.
\end{aligned}
\label{green_fns_fourier}
\eeq
Note that $\tilde\mg_0(-\nu)=\tilde\mg_0^*(\nu)$, $\mg_+(\nu)=\mg_-^*(-\nu)$, and that
$\tilde x(-\nu)=\tilde x^*(\nu)$ as expected because $x(t)$ is a real-valued
function.
\subsubsection{The NSD calculation}
In the same way as performed previously in Ref. ~\cite{batista2022gain}, 
we can write the NSD of $\tilde x(\nu)$, $\tilde x_\mathrm{L}(\nu)$, and
$\tilde y_\mathrm{L}(\nu)$, respectively,
 as
\beq
\resizebox{.92\hsize}{!}{$
\begin{aligned}
\label{S_Nnu_lo}
S_\mathrm{\tilde x}(\nu)&=\lim_{\Delta\nu\rightarrow 0^+}\displaystyle\int^{\nu+\Delta\nu}_{\nu-\Delta\nu} 
\dfrac{\langle \tilde x(-\nu)\tilde x(\nu')\rangle}{2\uppi}\; d\nu'
=2D\left[|\tilde\mg_0(\nu)|^2+|\mg_+(\nu)|^2+|\mg_-(\nu)|^2\right], \\
S_{\tilde x_\mathrm{L}}(\nu) &=\lim_{\Delta\nu\rightarrow 0^+}\displaystyle\int^{\nu+\Delta\nu}_{\nu-\Delta\nu} 
\dfrac{\langle \tilde x_\mathrm{L}(-\nu)\tilde x_\mathrm{L}(\nu')\rangle}{2\uppi}\; d\nu'
=
\frac{D}2\frac{\left|\tilde\mg_0(\nu-\omega)+\mg_+(\nu+\omega)
\right|^2+\left|\tilde\mg_0(\nu+\omega)+\mg_-(\nu-\omega)\right|^2}{1+\tau^2\nu^2},\\
S_{\tilde y_\mathrm{L}}(\nu) &=\lim_{\Delta\nu\rightarrow 0^+}\displaystyle\int^{\nu+\Delta\nu}_{\nu-\Delta\nu} 
\dfrac{\langle \tilde y_\mathrm{L}(-\nu)\tilde y_\mathrm{L}(\nu')\rangle}{2\uppi}\; d\nu'=
\frac{D}2\frac{\left|\tilde\mg_0(\nu-\omega)-\mg_+(\nu+\omega)
\right|^2+\left|\tilde\mg_0(\nu+\omega)-\mg_-(\nu-\omega)\right|^2}{1+\tau^2\nu^2}.
\end{aligned}
$}
\eeq
To obtain these results, we used the fact that the autocorrelation
of noise in frequency space is given by 
$\langle\tilde r(\nu)\tilde r(\nu')\rangle=4\uppi D\delta(\nu+\nu')$.
The functions $S_{\tilde x_{\mathrm{L}}}(\nu)$ and 
$S_{\tilde y_\mathrm{L}}(\nu)$ represent the stationary NSDs
of the LIA $X$ and $Y$ channel outputs in the frequency domain.
At $\nu=0$, these results are in agreement with the NSDs for 
the real and imaginary parts of $\tilde x(\omega)$. 
See the appendix \ref{sec:lock-in} and the discussion on squeezing further below.

The NSD of a simple harmonic oscillator with additive noise is given by
$S^\mathrm{ho}_{\tilde x}(\omega)=2D|\chi(\omega)|^2$.
Hence, we have 
\[
\frac1{2\uppi}\int_{-\infty}^\infty
S^\mathrm{ho}_{\tilde x}(\omega)d\omega=\frac{D}{\gamma\omega_0^2},
\]
when the noise is thermal, we have $D\propto T_0$ , where $T_0$ is the ambient
temperature.
Similarly, in a parametric resonator with feedback, we have
\[
\frac1{2\uppi}\int_{-\infty}^\infty
S_{\tilde x}(\omega)d\omega=\frac{\tilde D}{\gamma\omega_0^2},
\]
where $\tilde D\propto T$, the temperature of the parametric resonator.
As the constants of proportionality are the same, we define the
effective temperature as the ratio $\frac{\tilde
D}D=\frac T{T_0}=T_\mathrm{eff}$.
\subsubsection{The squeezing calculation}
Here, we investigate the fluctuations in the real and imaginary parts of
$\tilde x(\nu)$ ($\tilde x'(\nu)$ and $\tilde x''(\nu)$, respectively).
When $\nu\neq\omega$ the dispersions of these stochastic variables are the
same and there is no correlation between them \cite{batista2025deep}.
When $\nu=\omega$, the dispersions of $\tilde x'(\omega)$ and $\tilde
x''(\omega)$ become different from one another and, in general, a correlation
arises between them.
These results can be observed numerically and also experimentally.
We assume that the stochastic variates, $\tilde x'(\omega)$ and 
$\tilde x''(\omega)$,  are generated by a two-variate Gaussian distribution with
zero mean and, in general, with correlation.
Below, we show how to obtain analytical expressions for the dispersions in
$\tilde x'(\omega)$, $\tilde x''(\omega)$, and the correlation between them.

Based on Eq.~\eqref{x_nu_fb}, we find the same type of expressions for
fluctuations squeezing as in Eqs.~(29)-(30) of Ref.~\cite{batista2025deep}.
We can write the real and imaginary parts of $\tilde x(\nu)$, given in
Eq.~\eqref{x_nu_fb}, as
\beq
\begin{aligned}
    \tilde x'(\nu) &=\tilde\mg_0'(\nu)\tilde r'(\nu)-\tilde\mg_0''(\nu)\tilde r''(\nu)+\left[
    \mg_+'(\nu)\tilde r'(\nu-2\omega)-\mg_+''(\nu)\tilde r''(\nu-2\omega)
    \right.\\
    +&\left.
    \mg_+'(-\nu)\tilde r'(\nu+2\omega)+\mg_+''(-\nu)\tilde r''(\nu+2\omega)
    \right],\\
    \tilde x''(\nu) &=\tilde \mg_0'(\nu)\tilde r''(\nu)+\tilde \mg_0''(\nu)\tilde r'(\nu)+\left[
    \mg_+'(\nu)\tilde r''(\nu-2\omega)+\mg_+''(\nu)\tilde r'(\nu-2\omega)
    \right.\\
    +&\left.
    \mg_+'(-\nu)\tilde r''(\nu+2\omega)-\mg_+''(-\nu)\tilde r'(\nu+2\omega)
    \right].
\end{aligned}
\label{xr_xi}
\eeq
Here, we denote the real and imaginary part of a given complex function $f$ as
$f'$ and $f''$, respectively.
Using the parity properties of the Fourier transform of a real function
\beq
\begin{aligned}
    \tilde r'(\nu)&=\tilde r'(-\nu),\\
    \tilde r''(\nu) &=-\tilde r''(-\nu),
\end{aligned}
\eeq
and the following statistical averages of white noise in the frequency domain:
\beq
\begin{aligned}
\langle \tilde r'(\nu)\tilde r'(\nu')\rangle &=2\uppi
D\left[\delta(\nu-\nu')+\delta(\nu+\nu')\right],\\
\langle \tilde r'(\nu)\tilde r''(\nu')\rangle &=0,\\
\langle \tilde r''(\nu)\tilde r''(\nu')\rangle &=2\uppi
D\left[\delta(\nu-\nu')-\delta(\nu+\nu')\right],
\end{aligned}
\eeq
we obtain the dispersions in quadrature and the correlation between them of the
response of the parametric resonator to added noise, when $\nu=\omega$.
They are given by
\beq
\begin{aligned}
\sigma_\mathrm{c}^2(\omega)&=\lim_{\Delta\nu\rightarrow 0^+}\int^{\omega+\Delta\nu}_{\omega-\Delta\nu} 
\langle \tilde x'(\omega)\tilde x'(\nu')\rangle\; d\nu'\\
&\approx2\uppi D\left[|\tilde\mg_0(\omega)|^2+|\mg_+(\omega)|^2
+2\Re\left\{\tilde\mg_0(\omega)\mg_+(\omega)\right\}\right],\\
\sigma_\mathrm{s}^2(\omega)&=\lim_{\Delta\nu\rightarrow 0^+}\int^{\omega+\Delta\nu}_{\omega-\Delta\nu} 
\langle \tilde x''(\omega)\tilde x''(\nu')\rangle\; d\nu'
\\
&\approx2\uppi D\left[|\tilde\mg_0(\omega)|^2+|\mg_+(\omega)|^2
-2\Re\left\{\tilde\mg_0(\omega)\mg_+(\omega)\right\}\right],
\\
\sigma_{\mathrm{cs}}(\omega)&=\lim_{\Delta\nu\rightarrow 0^+}\int^{\omega+\Delta\nu}_{\omega-\Delta\nu} 
\langle \tilde x'(\omega)\tilde x''(\nu')\rangle\; d\nu'=
4\uppi D \Im\left\{\tilde\mg_0(\omega)\mg_+(\omega)\right\}.
\end{aligned}
\label{sigma2_om}
\eeq
At $\nu=\omega$, we have approximately
\beq
\begin{aligned}
\tilde\mg_0(\omega)&\approx
\frac{\alpha^*(\omega)\chi(\omega)}{|\alpha(\omega)|^2-|\beta(-\omega)|^2},\\
\mg_+(\omega)&\approx
-\frac{\beta^*(-\omega)\chi^*(\omega)}{|\alpha(\omega)|^2-|\beta(-\omega)|^2},\\
\Re\left\{\tilde\mg_0(\omega)\mg_+(\omega)\right\} &\approx
-\frac{\Re\left\{\alpha^*(\omega)\beta^*(-\omega)\right\}}
{\left(|\alpha(\omega)|^2-|\beta(-\omega)|^2\right)^2}|\chi(\omega)|^2,\\
\Im\left\{\tilde\mg_0(\omega)\mg_+(\omega)\right\} &\approx
-\frac{\Im\left\{\alpha^*(\omega)\beta^*(-\omega)\right\}}
{\left(|\alpha(\omega)|^2-|\beta(-\omega)|^2\right)^2}|\chi(\omega)|^2.
\end{aligned}
\eeq
After replacing these expressions in Eq.~\eqref{sigma2_om}, we obtain the dispersions and correlation to be given by
\beq
\begin{aligned}
\sigma_\mathrm{c}^2(\omega)&\approx
2\uppi D\frac{\left|\alpha^*(\omega)+\beta(-\omega)\right|^2}
{\left[|\alpha(\omega)|^2-|\beta(-\omega)|^2\right]^2}|\chi(\omega)|^2,\\
\sigma_\mathrm{s}^2(\omega)&\approx
2\uppi D\frac{\left|\alpha^*(\omega)-\beta(-\omega)\right|^2}
{\left[|\alpha(\omega)|^2-|\beta(-\omega)|^2\right]^2}|\chi(\omega)|^2,\\
\sigma_{\mathrm{cs}}(\omega)&\approx
-4\uppi D\frac{\Im\left\{\alpha^*(\omega)\beta^*(-\omega)\right\}}
{\left(|\alpha(\omega)|^2-|\beta(-\omega)|^2\right)^2}|\chi(\omega)|^2,
\end{aligned}
\label{eq:sigmas}
\eeq
where
\[
\begin{aligned}
\alpha(\omega) &=1-\frac{\eta\omega\tau\chi(\omega)}{1-2\imath\omega\tau},\\
\beta^*(-\omega) &= -\frac{\imath\chi(\omega)}2(F_\mathrm{p}+\eta).
\end{aligned}
\]
Note that if $F_\mathrm{p}=\eta=0$, we find
$\sigma^2_c=\sigma^2_s=\sigma_0^2=2\uppi D|\chi(\omega)|^2$ and
$\sigma_{\mathrm{cs}}=0$ as expected for a harmonic oscillator driven by white
noise.
Otherwise, for $F_\mathrm{p}\neq0$ and $\eta\neq0$, in the case $\omega=\omega_0$ ($\Omega=0$) 
we obtain
\[
\begin{aligned}
\alpha(\omega_0) &=1-\frac\eta{2\gamma\omega_0},\\
\beta^*(-\omega) &= \frac{F_\mathrm{p}+\eta}{2\gamma\omega_0}.
\end{aligned}
\]
Replacing these results in Eq.~\eqref{eq:sigmas}, we find
\beq
\begin{aligned}
\sigma_\mathrm{c}^2(\omega)&\approx
\frac{2\uppi D}{(\gamma\omega_0-F_\mathrm{p}/2)^2},\\
\sigma_\mathrm{s}^2(\omega)&\approx
\frac{2\uppi D}{(\gamma\omega_0+\eta+F_\mathrm{p}/2)^2},\\
\sigma_{\mathrm{cs}}(\omega)&\approx0.
\end{aligned}
\label{eq:sigmas2}
\eeq
We find that the standard deviations $\sigma_\mathrm{c}$ and $\sigma_\mathrm{s}$ are proportional
to the maximum and minimum gains $G_{\mathrm{avg}}^{\mathrm{max}}$ and
$G_{\mathrm{avg}}^{\mathrm{min}}$,
respectively, given in Eq.~\eqref{G_minG_max}.
This is expected since the parametric amplifier responds to an added ac
drive or to noise in the same way, since it is a linear amplifier.
\subsection{FT approach}
\label{Floquet}
We now apply the same FT method developed previously in
Refs.~\cite{batista2024amplification,batista2025deep} to
obtain the average fluctuations (the NSDs and the squeezing) of the parametric
resonator with LIA feedback. 
This became possible because we transformed the original integro-differential
model of Eq.~\eqref{pa_lockin_feedback_amp} into the ODE system with additive
noise of Eq.~\eqref{po_lockin_feedback_noise}.
The FT method has one important advantage over the approximate Fourier transform
method we developed in the previous section: we can find
far more precise estimates for the NSD and for the squeezing over a wider range
of parameters.
\subsubsection{The NSD calculation}
From Eq.~\eqref{fundEqX_iF_j}, we find that the parametric resonator response to noise is given by
\beq
\tilde x(\nu)=\tilde G_{0,01}(\nu)\tilde r(\nu)
+G_{+, 01}(\nu)\tilde r(\nu-2\omega)
+G^*_{+,01}(-\nu)\tilde r(\nu+2\omega),
\eeq
where the additive noise vector is given by
\beq
F(t)=
\left(
\bea{c}
0\\
r(t)\\
0
\ea
\right).
\eeq
Hence, we find that the NSD for the SDOF parametric resonator with LIA feedback
and additive noise is given by
\beq
S_\mathrm{\tilde x}(\nu)=2D\left(\left|\tilde G_{0,01}(\nu)\right|^2+\left|G_{+,01}(\nu)\right|^2+\left|G_{-,01}(\nu)\right|^2\right).
\label{S_Npar_fb}
\eeq
The Green's functions in the frequency domain are obtained from applying FT to 
Eq.~\eqref{po_lockin_feedback_noise} and using the techniques developed
in Ref.~\cite{batista2024amplification}. 
\subsubsection{The squeezing calculation}
\label{sec:squeezingN}
The calculation of the dispersions in $\tilde x'(\nu)$ and $\tilde x''(\nu)$ and
the correlation between them is performed in the same way as in
Eqs.~\eqref{sigma2_om}, but with $\tilde\mg_0(\omega)$ and $\mg_+(\pm\omega)$
replaced by $\tilde G_{0,01}(\omega)$ and $G_{+,01}(\pm\omega)$, respectively.
\FloatBarrier
\section{Numerical results and discussion}
\label{results}
\FloatBarrier
Here, we discuss the numerical results of the parametric resonator and amplifier
with LIA feedback.
Our objective is to verify numerically the main theoretical predictions of our
model.
We start by investigating only the deterministic equations of motion, since
they are simpler to study, the integration algorithms are more precise,
and they are a lot lighter computationally than the integration of stochastic
differential equations.
Also, conceptually, the theoretical methods used to investigate stochastic
differential equations are more complex and less well developed than those
of the deterministic equations.
Additionally, understanding the types of bifurcations to
parametric instability that occur is a necessary condition for the study of
fluctuations in these systems, as the strongest squeezing and heating occur in
the parameter space regions closest to the instability thresholds and cooling is
strongest in regions far from these thresholds.
Simple ringdown experiments help detect the onset of HBs or SNBs in these
systems.
Moreover, features of the FEs will appear in the NSDs.
Furthermore, as the amplifier is linear in its response to an added signal or
noise, understanding the amplification gain of the parametric amplifier with LIA
feedback to an ac signal gives us insight into the response to noise.
We can predict whether squeezing, heating, or cooling will occur.

\begin{figure}[!ht]
\centerline{\includegraphics[{scale=0.6}]{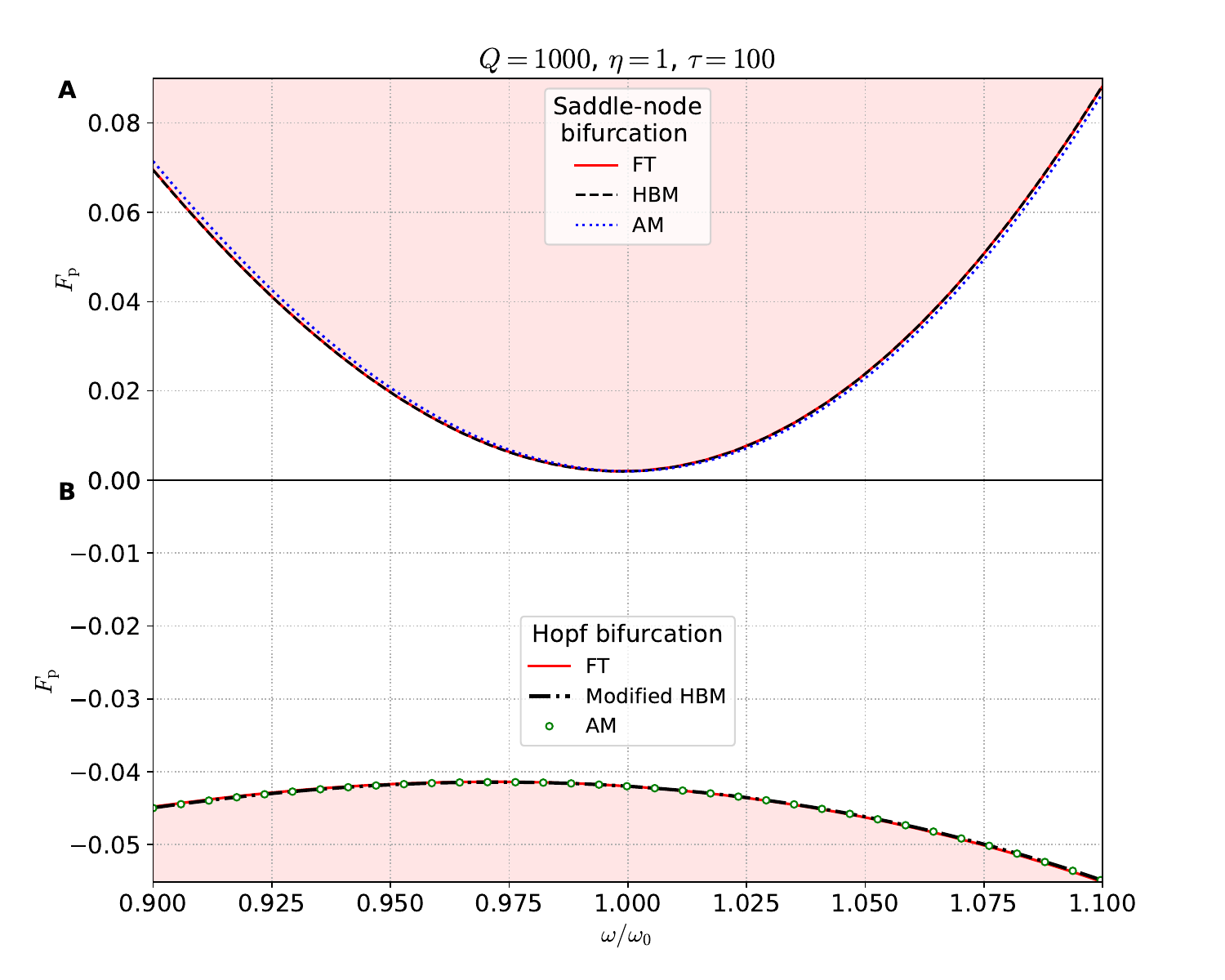}}
\caption{Parametric instability threshold lines for the SDOF parametric resonator
with LIA feedback as described by Eq.~\eqref{pa_lockin_feedback_amp2} with
$F_\mathrm{s}=0$. 
\textbf{\textsf{A}} The threshold line to instability occurs at a
SNB (when one real FM becomes equal to 1).
Comparison between the FT prediction, the HBM prediction (from the
equation $\det M=0$ in Eq.~\eqref{eq:detM}), and the averaging
prediction (the positive root of Eq.~\eqref{eq:threshold_avg}).
\textbf{\textsf{B}} The threshold line to instability occurs at a HB (when the
module of the complex conjugate pair of FMs becomes equal to 1).
Very good agreement is obtained between the FT prediction, the HBM prediction,
which is obtained from solving the algebraic system from
Eq.~\eqref{Fp_delta_system}, and the AM prediction \eqref{eq:Hopf_line_avg} for
the HB line. 
}
\label{fig:threshold_FB}
\end{figure}

In Fig.~\ref{fig:threshold_FB} we show the instability threshold lines
of the parametric resonator with LIA feedback as described by Eq.~\eqref{pa_lockin_feedback_amp2}.
In panel \textbf{\textsf{A}}, approximations to the threshold lines (via a SNB)
are obtained approximately via the AM or the HBM, or more exactly via the FT
method.
The dotted blue line is obtained by the AM, positive branch of
Eq.~\eqref{eq:threshold_avg}, while the dashed line is obtained with the HBM,
Eq.~\eqref{eq:detM}.  The FT threshold line occurs when one real FM reaches
module 1.
One can see an excellent agreement between the HBM predictions with FT results
for the instability line. 
The averaging approximation is also good, but it is slightly off for increasing
detuning. 
In panel \textbf{\textsf{B}}, approximations to the HB threshold lines are
obtained approximately with the AM, with the modified HBM, or more exactly with
the FT.
The AM threshold line occurs when there is a complex conjugate pair of imaginary
eigenvalues of the Jacobian matrix of the ODE system in Eq.~\eqref{uvz_avg}.
The analytical expression for this line is given by the negative branch of
Eq.~\eqref{eq:Hopf_line_avg}.
Using the modified HBM, a HB occurs at a quasi-periodic solution
as given in Eq.~\eqref{eq:quasi_periodic}.
The dashed threshold line is obtained from the numerical solution of
Eq.~\eqref{Fp_delta_system}.
The FT threshold occurs when a complex conjugate pair of FMs reach module 1.
The AM and HBM lines agree very well with the FT prediction for the HB
to instability.
 
\begin{figure}[!ht]
\centerline{\includegraphics[{scale=0.7}]{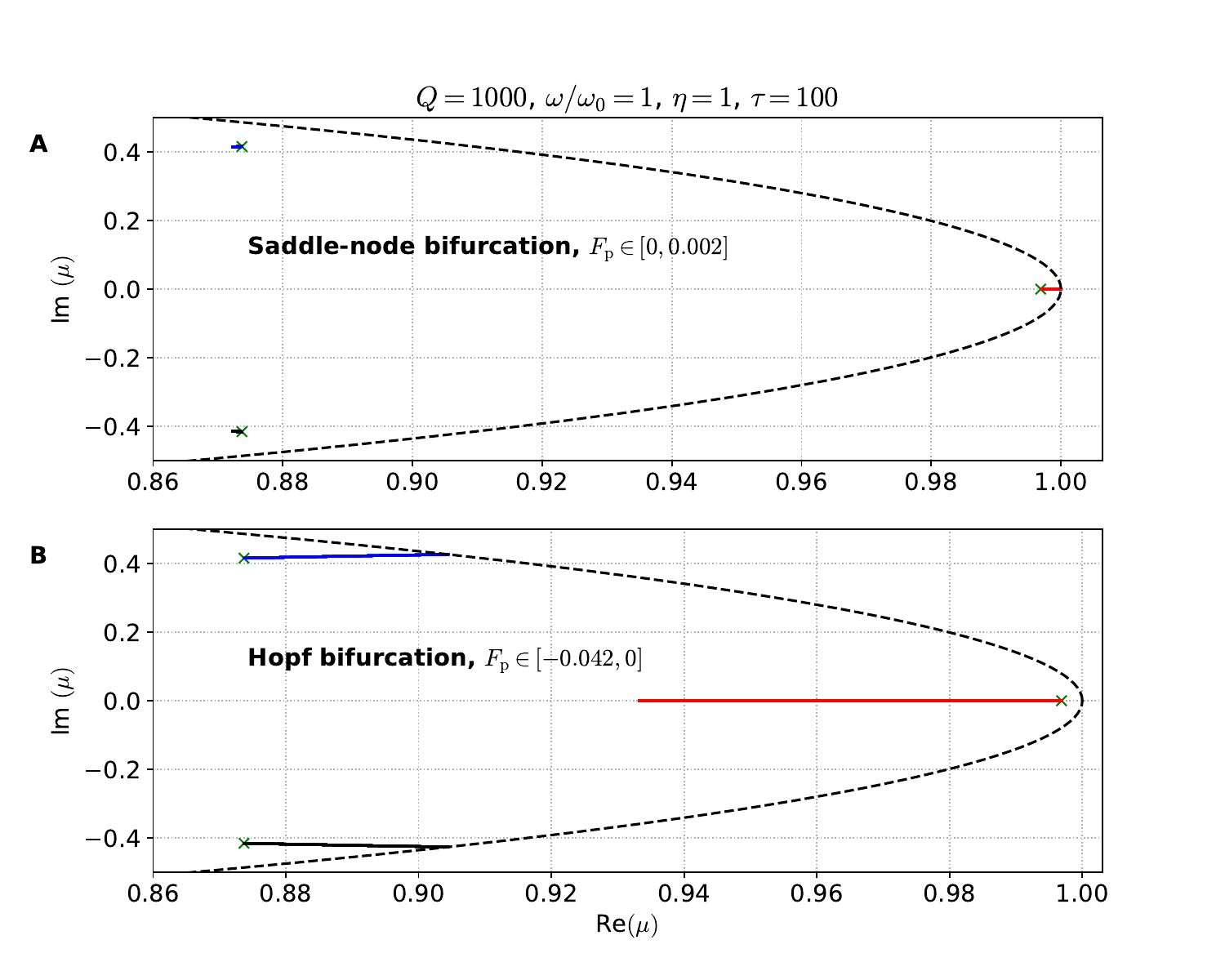}}
\caption{Plots of the FMs as the pump amplitude is varied
for the parameters shown at the top of the figure.
The vertical axis was compressed for better visibility.
The dashed line corresponds to the instability threshold at $|\mu|=1$.
The symbols '{\color{green} $\times$}' represent the FMs at $F_\mathrm{p}=0$.
\textbf{\textsf{A}} The pump amplitude $F_\mathrm{p}$ is varied from $0$ to $0.002$. 
A SNB occurs at $F_\mathrm{p}\approx0.002$ when the real FM
reaches the values $1$ while the complex FMs still have magnitude less than 1.
\textbf{\textsf{B}} The pump amplitude $F_\mathrm{p}$ is varied from $0$ to $-0.042$.
A HB occurs at $F_\mathrm{p}\approx -0.042$ when the pair of complex FMs
reaches the unit circle while the real FM still has magnitude less than 1.
}
\label{fig:multipliers}
\end{figure}

In Fig.~\ref{fig:multipliers} \textbf{\textsf{A}} we plot the FMs
in the complex plane as the pump amplitude $F_\mathrm{p}$ is varied from 0 to
$0.002$.
A SNB occurs at $F_\mathrm{p}\approx0.002$ when the real FM
reaches the value $1$ while the complex FMs still have magnitude less than 1.
In panel \textbf{\textsf{B}} we plot the FMs in the complex
plane as $F_\mathrm{p}$ is varied from 0 to $-0.042$ (there is a phase of $180^o$ in the
pump relative to the pump in panel \textbf{\textsf{A}}). 
For $F_\mathrm{p}=-0.042$, the complex conjugate pair of FMs reaches module 1
while the real FM has module less than 1.
This is a hallmark of a HB. 
This bifurcation is only possible because the dynamical system of
Eq.~\eqref{pa_lockin_feedback_amp2} has three FMs and also because
the product of these multipliers, according to Liouville's formula, is
$\mu_1\mu_2\mu_3=\ee^{-2\uppi(\gamma+\tau^{-1})/\omega}<1$ for any set of
parameters.
Hence, when only one FM is real ($\mu_r$) and the other two FMs
are the complex conjugate pair $\mu_c$ and $\mu_c^*$,
a SNB occurs when $\mu_r=1$ and $|\mu_c|<1$, while at a
HB: $|\mu_r|<1$ and $|\mu_c|=1$. 

\begin{figure}[!ht]
\centerline{\includegraphics[{scale=0.7}]{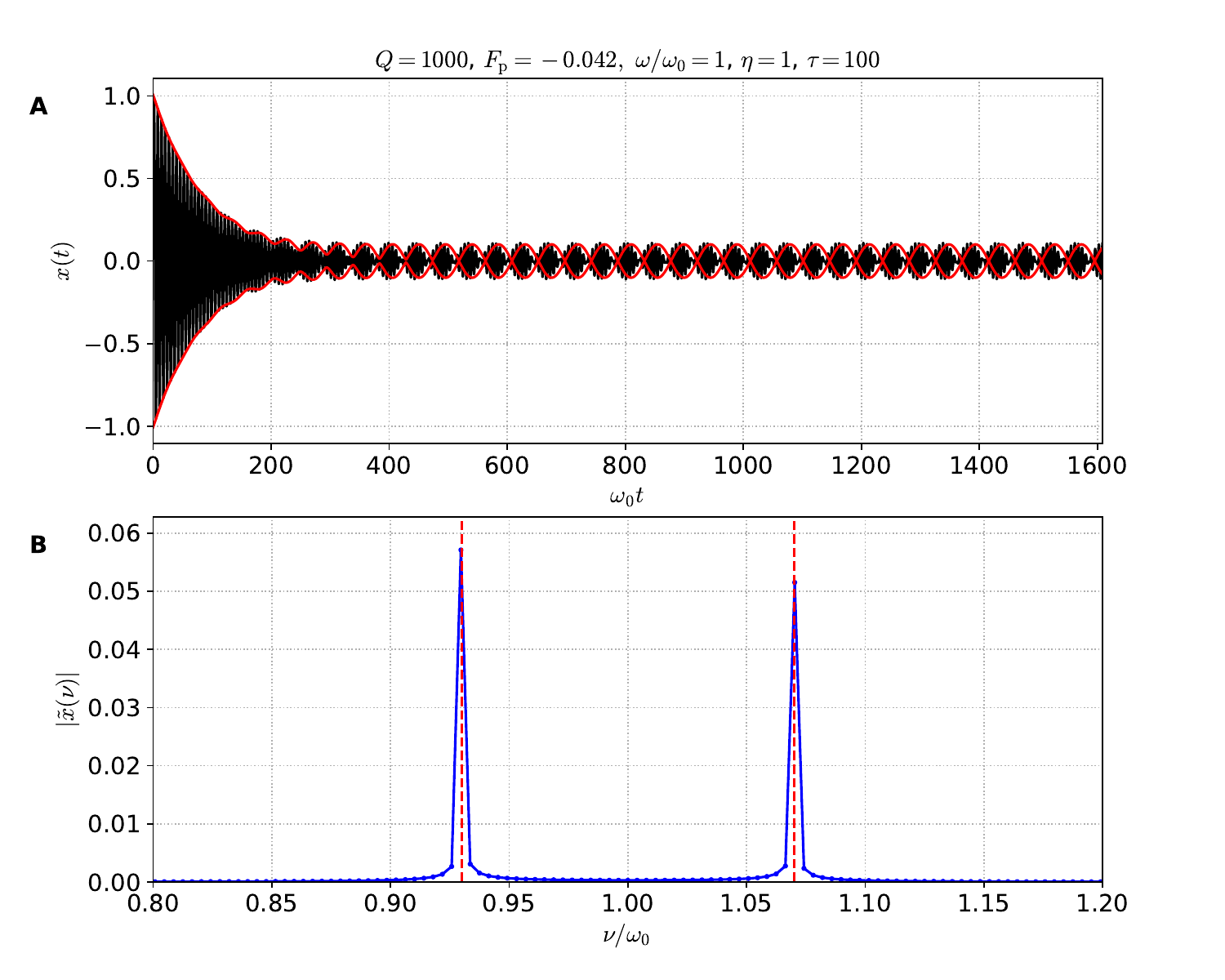}}
\caption{
\textbf{\textsf{A}}
A time-series transient of $x(t)$ obtained from numerical integration of
Eqs.~\eqref{pa_lockin_feedback_amp2} with parameters set near a HB point. The envelopes (red lines) are obtained
from the averaging approximation by integrating Eqs.~\eqref{uvz_avg}.
The envelopes are given by $\pm \sqrt{u^2(t)+v^2(t)}$.
\textbf{\textsf{B}}
Fourier transform of the stationary part of the time series shown in panel
\textbf{\textsf{A}}.
The two peaks are a signature of quasi-periodic behavior that arises near the
onset of a HB.
The vertical dashed lines were obtained from the HBM by solving
for $\Delta$ in Eq.~\eqref{Fp_delta_system}.
These peaks can also be found with the AM or with FT. 
}
\label{fig:transient}
\end{figure}

In Fig.~\ref{fig:transient} \textbf{\textsf{A}} we show a transient of $x(t)$ 
obtained from numerical integration. 
The parameters are set near a HB.
One sees that the AM provides a very good envelope to the initially decaying
oscillations and a fairly good approximation to the cyclo-stationary
oscillations as they emerge after the transient dies out near a HB, except for
the slight detuning that keeps increasing with time.
We notice that care should be taken to avoid choosing the initial conditions of
the ODE system \eqref{uvz_avg} on its stable manifold. 
In this case, one could be led to erroneously think that the AM fails to capture
the HB.
In panel \textbf{\textsf{B}} we plot the Fourier transform of the stationary
part of the time series shown in panel \textbf{\textsf{A}}.
The two peaks are a signature of the quasi-periodic behavior that
arises near the onset of a HB.
The dashed lines were obtained from the AM using Eq.~\eqref{eq:Delta_avg}.
Alternatively, it can also be obtained from HBM by solving
Eq.~\eqref{Fp_delta_system} or, more precisely, from FT, where $\Delta$ is the
imaginary part of a complex FE.

\begin{figure}[!ht]
\centerline{\includegraphics[{scale=0.7}]{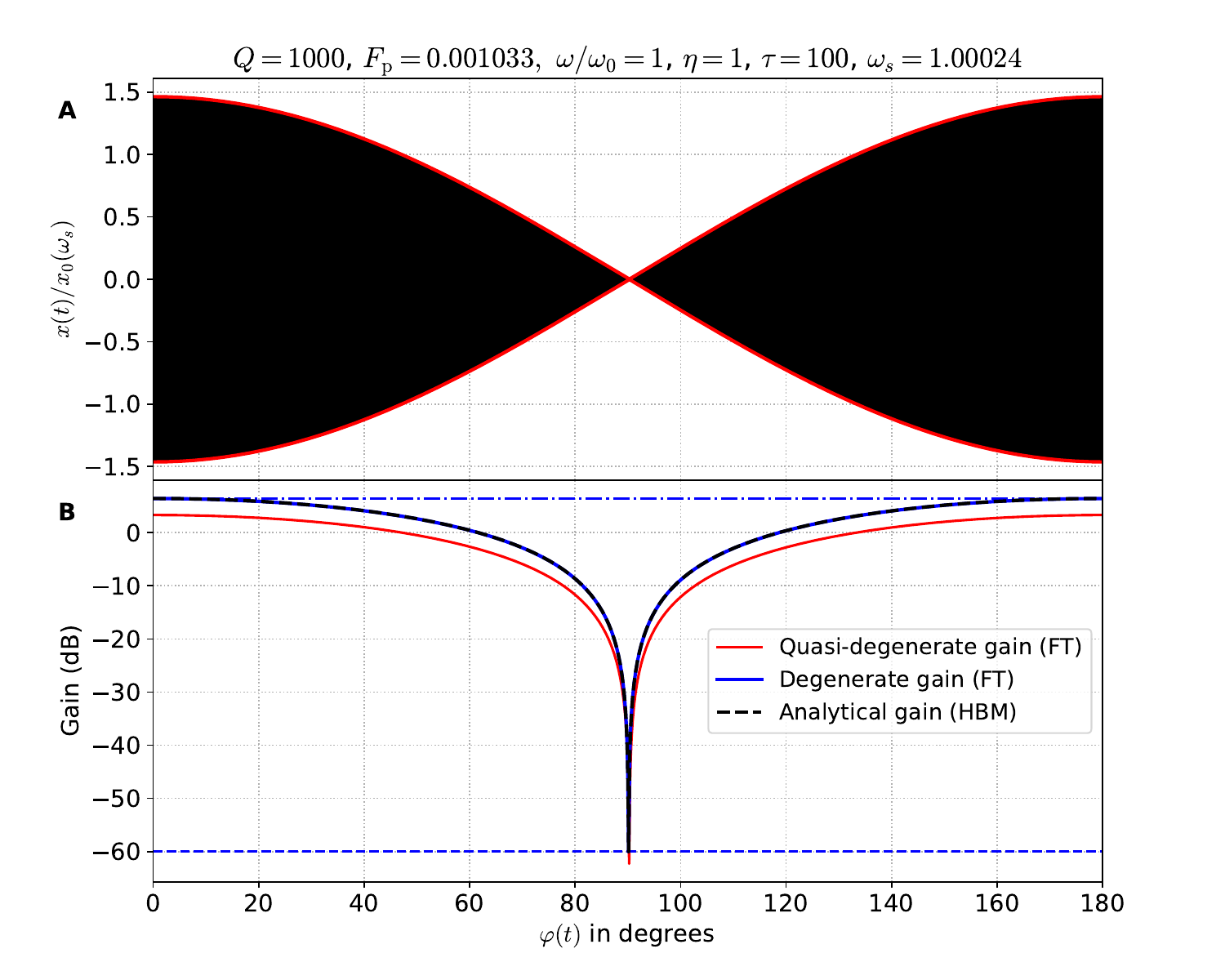}}
\caption{Gain as a function of phase for the parametric amplifier with LIA
feedback whose dynamics is given in Eq.~\eqref{pa_lockin_feedback_amp2} with
parameters set near the onset of a SNB. 
\textbf{\textsf{A}} We plot the normalized cyclo-stationary response $x(t)/x_0$ as a
function of phase $\varphi$, where $x_0$ is the amplitude of oscillations of
the forced harmonic resonator (when $F_\mathrm{p}=\eta=0$).
There is a slight detuning between $\omega$ and $\omega_\mathrm{s}$ so that the phase is
swept very slowly, almost quasi-statically.
The envelopes are obtained from Eq.~\eqref{eq:envelopes}, which are based on FT.  
\textbf{\textsf{B}} We use the positive envelope from \textbf{\textsf{A}} to obtain the
quasidegenerate gain in decibels.
The FT degenerate gain is obtained in Eq.~\eqref{gain_FT_fb}.
The analytical HBM gain is given in Eq.~\eqref{gain_HBM} with the help of 
Eq.~\eqref{eq:A_1}. The horizontal lines are the HBM approximations for
minimum and maximum gains obtained from Eq.~\eqref{G_minG_max}.
}
\label{fig:gain_FB}
\end{figure}

In Fig.~\ref{fig:gain_FB} \textbf{\textsf{A}}, we show a normalized
cyclo-stationary response to the ac excitation at angular frequency $\omega_\mathrm{s}$
which is slightly detuned from half the pump frequency $\omega$.
The envelopes obtained from FT provide an excellent fit for the oscillations.
The parameters are set near the onset of a SNB.
In panel \textbf{\textsf{B}}, we achieve very deep deamplification around
$-60~\si{dB}$, which is
much less than the $-6~\si{dB}$ lower limit of standard classical parametric deamplification.
This shows that deep squeezing can be obtained in the SDOF parametric resonator
with LIA feedback.

\begin{figure}[!ht]
\centerline{\includegraphics[{scale=0.7}]{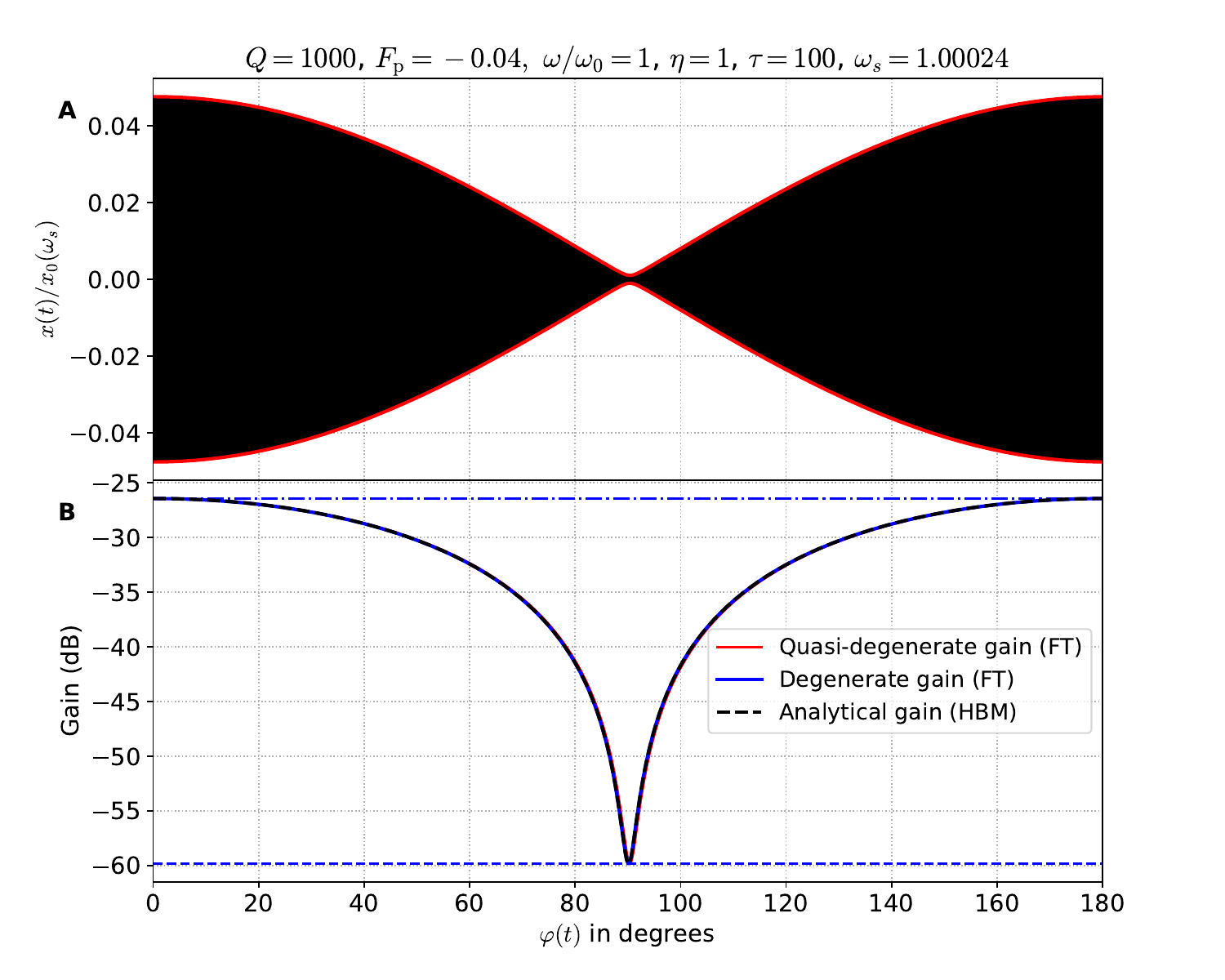}}
\caption{Gain as a function of phase for the SDOF parametric
amplifier with LIA feedback whose dynamics is given in
Eq.~\eqref{pa_lockin_feedback_amp2} with parameters set near the onset of a 
HB.
\textbf{\textsf{A}} The response of the SDOF parametric resonator with
feedback is far reduced compared with the amplitude of the harmonic oscillator
response.  
\textbf{\textsf{B}} There is attenuation in all phases.
This indicates that with these parameters the parametric resonator with
the proposed feedback scheme leads to cooling. 
}
\label{fig:gain_FB2}
\end{figure}

In Fig.~\ref{fig:gain_FB2} \textbf{\textsf{A}}, we again show a normalized
cyclo-stationary response to the ac excitation at angular frequency $\omega_\mathrm{s}$
which is slightly detuned from half the pump frequency $\omega$.
The response of the SDOF parametric resonator with
feedback is far reduced compared with the amplitude of the harmonic oscillator
response.  
The parameters are set near the onset of a HB.
In panel \textbf{\textsf{B}} we plot the gain as a function of phase for the
SDOF parametric amplifier with LIA feedback as described in
Eq.~\eqref{pa_lockin_feedback_amp2}.
There is attenuation in all phases.
This indicates that the parametric resonator with feedback set with these chosen
parameters should experience cooling when excited with an additive thermal
noise. 
Perhaps, the easiest way to understand why this happens comes from the
inspection of the FT Green's functions given in Eq.~\eqref{G_0G_+}.
The real FE is negative and not close to zero.
Hence, the resonance peak near $\nu=\omega$ is deamplified and
there is no gain in any phase.

\begin{figure}[!ht]
\centerline{\includegraphics[{scale=0.65}]{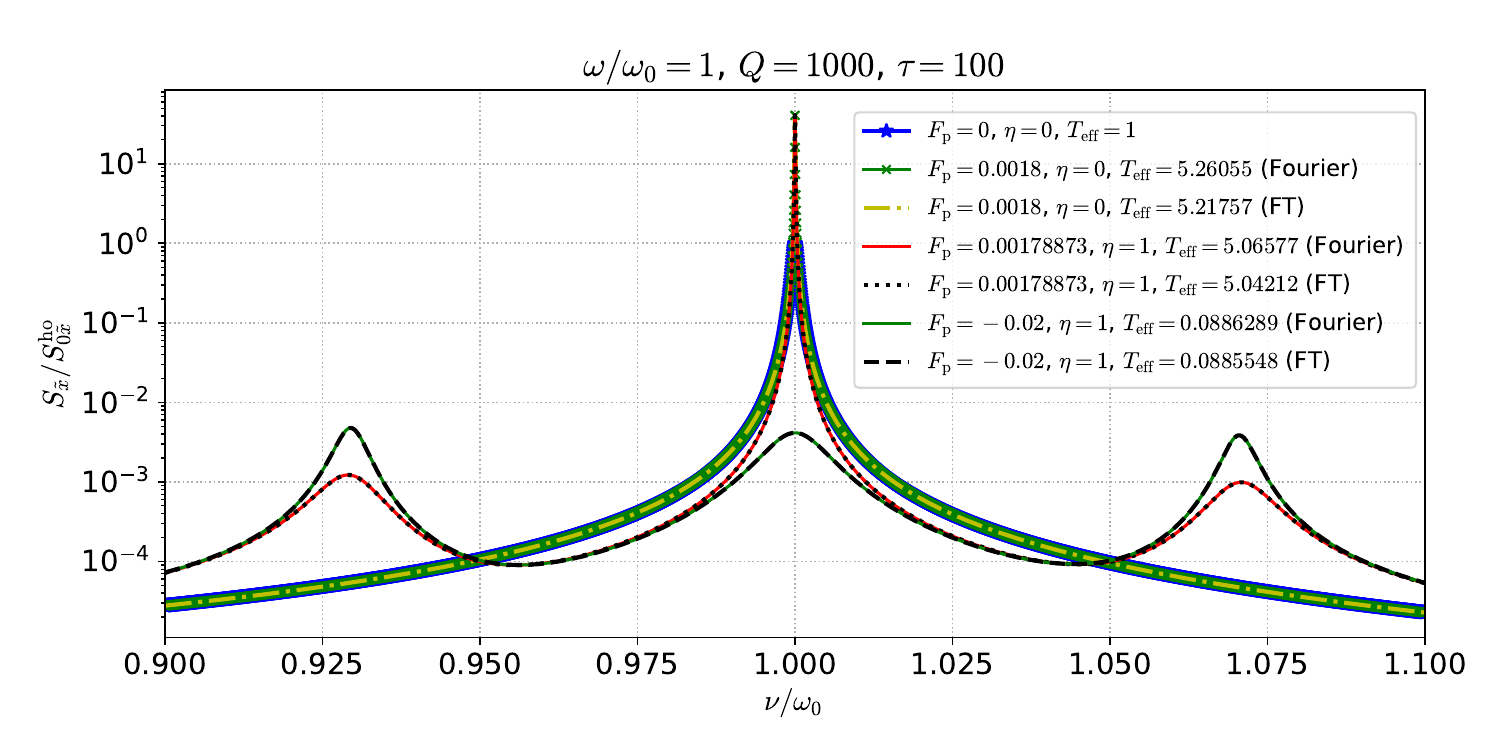}}
\caption{NSDs in logarithmic scale:
the harmonic oscillator (solid blue line), the SDOF parametric resonator using
Fourier analysis (solid black line) and using FT (dot dashed line), the SDOF
parametric resonator with LIA feedback (solid red line, using Fourier analysis,
and dotted line, using FT), with negative pump (solid green line (Fourier
analysis) and dashed black line (FT)).
$S_{0\tilde x}^{\mathrm{ho}}$ is the peak value of the harmonic oscillator NSD.
The Fourier analysis results are based on the analytical expression for the NSD
given in Eq.~\eqref{S_Nnu_lo}  and the FT results are based on the NSD given in
Eq.~\eqref{S_Npar_fb}.
}
\label{fig:NSD_FB}
\end{figure}

In Fig.~\ref{fig:NSD_FB}, we show various scaled NSDs: 
the harmonic oscillator (the blue solid line), parametric resonator without
feedback (the solid green line with the symbol $\times$).
This NSD was obtained from Fourier analysis, whereas the yellow dot-dashed line was obtained from FT. 
The SDOF parametric resonator with LIA feedback: positive pump (the solid red
line, using Fourier analysis, and dotted black line, using FT), negative pump
(solid green line (Fourier analysis) and dashed black line (FT)).
The FT and Fourier analysis yield basically the same results for the NSDs in the
dB scale.
We find that the effective temperature of the resonator with the parameters $F_\mathrm{p}=-0.02$ and $\eta=1$ is roughly
$0.089$ of the effective temperature of the harmonic oscillator in thermal
equilibrium, which we assume to be equal to 1.
The two sideband peaks in some of the NSDs are due to the complex conjugate pair
of Floquet exponents \cite{batista2024amplification}.
The peaks occur at $\omega\pm\Delta$, where $\Delta=\Im\rho$, where $\rho$ is
a complex Floquet exponent with positive imaginary part.
Alternatively, this is approximately the $\Delta$ obtained in
Eq.~\eqref{eq:Delta_avg} or in Eq.~\eqref{Fp_delta_system}.
If the pump becomes more negative the sideband peaks grow and the resonator 
warms again as one nears the HB line.

\begin{figure}[!ht]
\centerline{\includegraphics[{scale=0.65}]{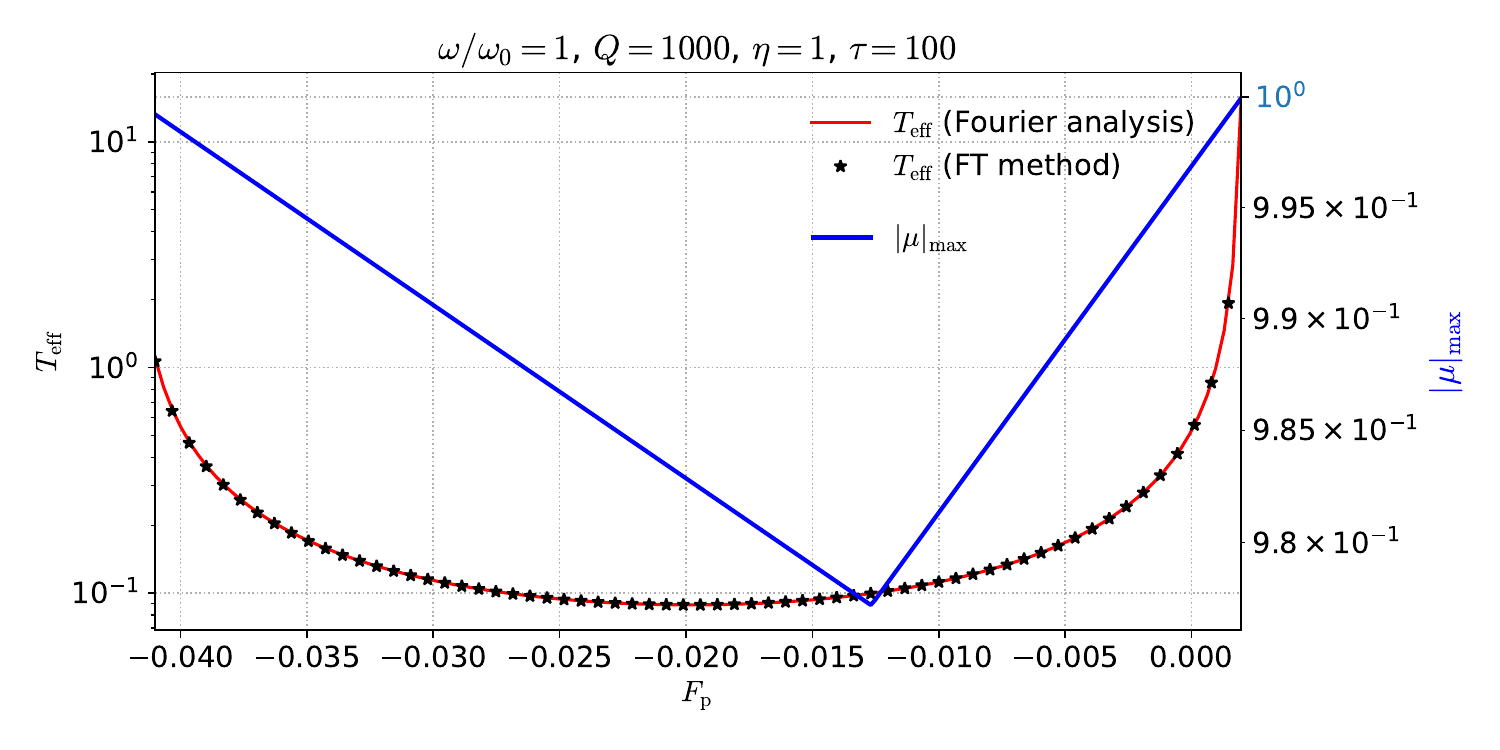}}
    \caption{ 
Effective temperature as a function of pump $F_\mathrm{p}$ from the threshold of
a HB (on the left) to the threshold of a SNB (on the right).
Negative $F_\mathrm{p}$ indicates there is an added $\uppi$ in the pump phase
relative to feedback phase in Eq.~\eqref{pa_lockin_feedback_amp2}. 
The blue line indicates the module of the largest FM whose
values can be seen on the right $y$ axis of the figure.
The smallest value of this curve corresponds to all FMs having the same module.
The Fourier analysis results for $T_{\mathrm{eff}}$ are based on the numerical
integration of the NSD $S_{\tilde x}$ given in Eq.~\eqref{S_Nnu_lo}  and the FT results for
$T_{\mathrm{eff}}$ are based on the numerical integration of the NSD given in
Eq.~\eqref{S_Npar_fb}.
}
\label{fig:eff_temps}
\end{figure}

In Fig.~\ref{fig:eff_temps}, we plot the effective temperature as a function of $F_\mathrm{p}$.
The minimum value of the largest FM in amplitude, $|\mu_{\mathrm{max}}|$, is given by
$\ee^{-2\uppi(\gamma+\tau^{-1})/(3\omega)}$ from Liouville's formula.
This occurs when all multipliers have the same module.
While this does not correspond to the minimum value of the effective
temperature, it provides an approximate estimate for pump value.
Based on the above expression, $\tau>0$ should be as small as possible for
smaller effective temperatures, but it cannot be zero, otherwise, the dynamical
system dimensions reduces to 2.
There should be an optimum $\tau$.
If the parametric resonator is in a heat bath, then $\gamma\propto D$, hence
by increasing $\gamma$ the noise level also increases, hence increasing $\gamma$
is not a good strategy for lowering the effective temperature.
The minimum temperature reached should correspond to the most efficient
cooling.

\begin{figure}[!ht]
\centerline{\includegraphics[{scale=0.45}]{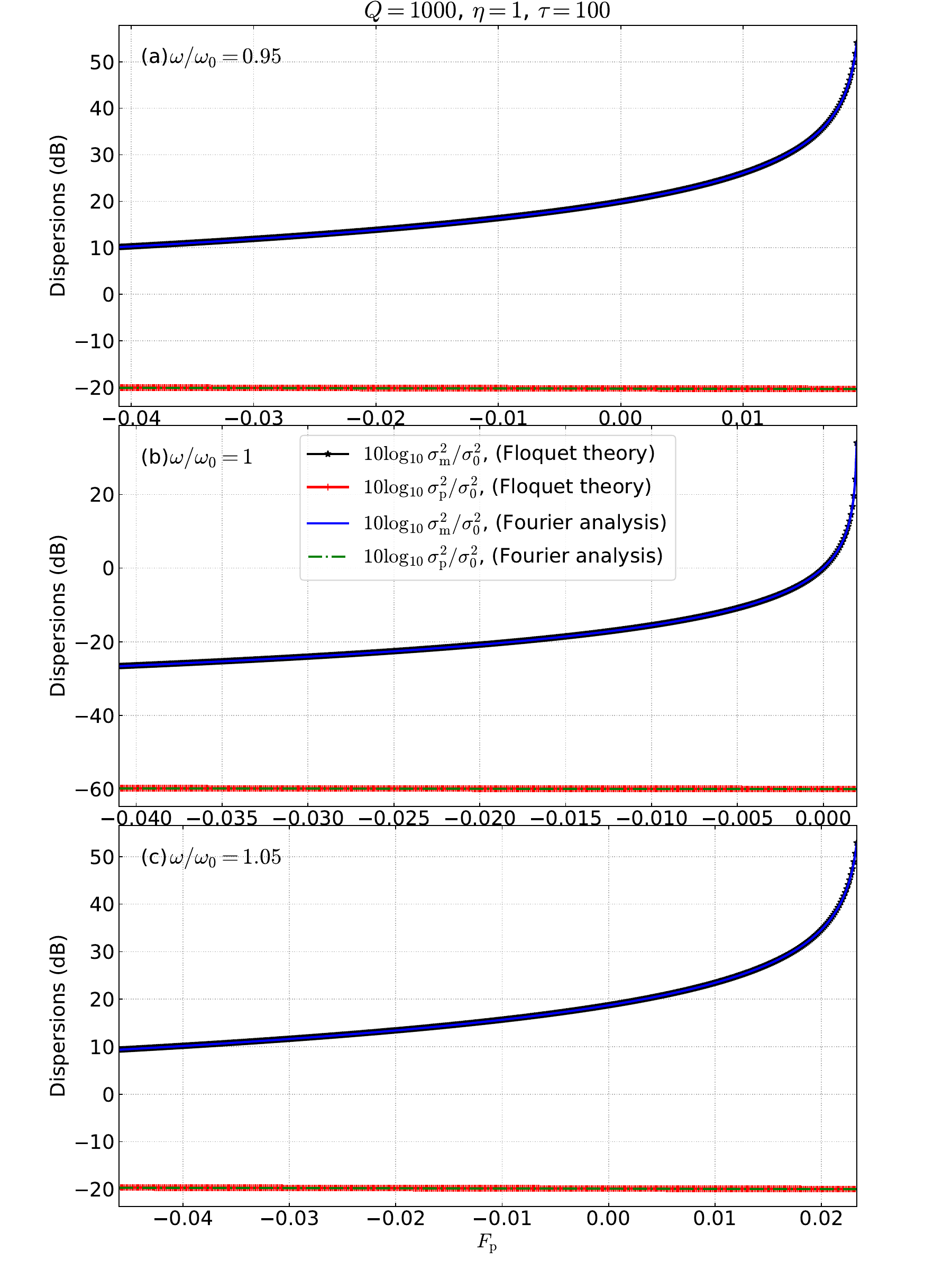}}
    \caption{ 
Fourier analysis and FT calculations of the diagonalized dispersions $\sigma_-$
and $\sigma_+$ in decibels relative to the equilibrium values with zero pump in
decibel scale in the SDOF parametric resonator with LIA feedback.
$\sigma_0$ is the standard deviation of the harmonic oscillator driven by noise.
The Fourier analysis results are based on Eqs.~\eqref{sigma2_om} where all
approximate Green's functions are given in Eq.~\eqref{green_fns_fourier}.
In the FT results, the Green's functions are given in Eqs.\eqref{G_0G_+}.
In obtaining all Green's functions, the feedback was taken into account.
Each plot ends very close to the instability thresholds. 
The common parameters used in all panels are given on top of the figure. 
}
\label{fig:squeezing_FB}
\end{figure}

In Fig.~\ref{fig:squeezing_FB}, we can see that very deep squeezing is
achieved for various values of $\omega$.
One sees very little dependence of the smallest dispersion on $F_\mathrm{p}$, whereas the
larger dispersion grows continuously from the smallest value near the HB until
it diverges at the SNB. 
When $\omega=\omega_0$, the squeezing is strongest and there is also
cooling near the HB.
These phenomena become weaker as detuning increases.
The derivation of the diagonalized standard deviations $\sigma_-$
and $\sigma_+$ in the
squeezing phenomenon is described in the appendix of
Ref.~\cite{batista2025deep}.
These standard deviations are obtained from the diagonalization of
bilinear form in the exponent of the Gaussian bivariate
probability distribution with dispersions and correlation
given in Eq.~\eqref{eq:sigmas}.
We obtained these results independently using the Fourier analysis and FT.
One can see a perfect agreement between them.

In this section, we verified that the main predictions of the theoretical
analysis of our models (deterministic and stochastic) are very accurate when
compared with corresponding numerical integration results.
The analytical results of the AM could be used to guide the experimentalist
in the search of interesting parameter space that could lead to deep squeezing
or cooling.
Its validity is largely limited though to small values of the parameters (in the
present case, of small $\Omega$, $\gamma$, $F_p$, $\eta$, $\tau^{-1}$). 
To increase cooling one has to reduce the value of $\tau$, but in doing so
the AM results become less accurate. 
In this case, one could still use HBM results on instability thresholds and gain. 
For more accurate results, one has to use FT, which is valid in all parameter
space.
The downside of the FT method is that it is far less analytical than the AM or 
the HBM.
This shows that the three methods we developed for the analysis of the
deterministic equations of motion of the parametric resonator or
amplifier with LIA feedback complement each other.
For the analysis of fluctuations in our resonator with additive noise, we have
developed the Fourier method (more analytical, but less precise) and we have
applied the previously developed FT method with Green's functions
(less analytical, but more precise) to investigate the phenomena of
thermomechanical noise squeezing and cooling.
The results thus obtained agreed with one another very well, what reinforces
our predictions.



\FloatBarrier
\section{Conclusion}
\label{conclusion}
Here, we have theoretically shown that incorporating feedback into a
parametric resonator fundamentally modifies its dynamics. 
By embedding the feedback into the equations of motion, we were able to
apply the same theoretical methods developed in Refs.
\cite{batista2024amplification, batista2025deep} to analyze its
behavior.
Due to the feedback scheme, the usual period-doubling
bifurcation of SDOF parametric resonators to instability was replaced by a SNB.
Furthermore, the feedback increased the dimension of the dynamical system,
what allowed the occurrence of HBs.
This opened two distinct ways of controlling fluctuations in
the resonator.
Our analysis, validated through multiple complementary methods (Fourier
analysis, FT, and direct numerical integration of the stochastic
differential equations involved) reveals that strong fluctuation suppression can
be achieved in two qualitatively different regimes.
In one of them, by employing a LIA feedback loop, the system's
response is no longer limited to the standard classical parametric squeezing picture (with
the $-6~\si{dB}$ lower bound \cite{rugar91}). 
While this is not a new result \cite{vinante2013feedback}, our analysis sheds
more light on it.
Near a SNB, the system exhibits deep squeezing in
one of its quadrature fluctuations while it strongly increases the fluctuations
in the other quadrature. 

The main advance and novelty of this paper is that we analyzed the full impact
of the LIA feedback-loop on the response of the parametric resonator to an ac
drive or to additive noise.
Near a HB, the feedback induces a regime of strong deamplification
and cooling, reducing the oscillator's effective temperature. 
Simple ringdown experiments could detect the onset of HBs in these
systems, while in the frequency domain, the calculation of the NSD
provides experimentally accessible signatures of each regime (one main
central peak near a SNB and three peaks with strong
sidebands near a HB). 
Furthermore, we show that there is an optimal amount of parametric pumping and
feedback to reach the strongest cooling.
Additionally, we also made estimates on the amount of squeezing that can be
achieved in our model.

The theoretical framework presented here is general and could be applied to a
wide range of physical implementations, from nano-mechanical resonators and
superconducting circuits to trapped ions and cold atoms. The two
bifurcations identified provide a clear experimental target: parameters can be chosen to either cool the system towards its quantum ground state
or to generate squeezed states necessary for back-action-evading measurements
\cite{schnabel2017squeezed}. Furthermore, extending this analysis to the quantum
regime presents a stimulating, yet challenging,  direction
for future research. 
One possibility is to apply the methods developed here to reach near
quantum-limited amplification in cooled optomechanical parametric resonators
\cite{bothner2020cavity}.
Additionally, our methods could be applied to reduce  phase-flip
errors in phase states of Josephson (or Kerr) parametric oscillators in the
presence of noise, an essential step for the realization of qubit quantum gates
in these systems \cite{grimm2020stabilization,bhai2023noise}.

\appendix
\section{The LIA operation}
\label{sec:lock-in}
\begin{figure}[!htb]
    \centerline{\includegraphics[scale=1.0]{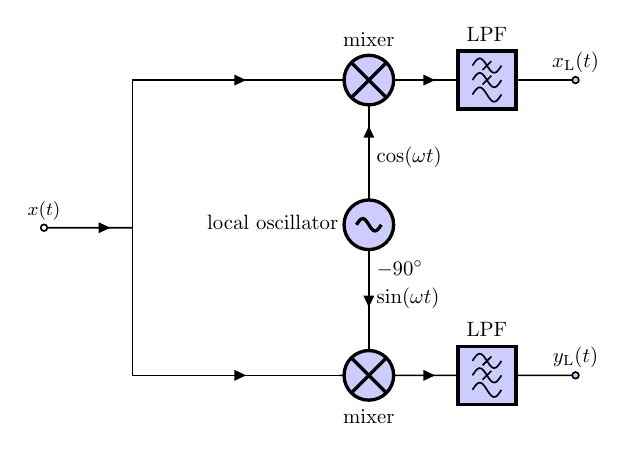}}
    \caption{
    The two-phase LIA block diagram. The low pass filters are $RC$ filters with
the same time constant. The mixers are multipliers.
    }
    \label{fig:lock-in}
\end{figure}

The two-phase LIA is essential for observing and measuring
thermo-mechanical squeezing \cite{rugar91}. 
Therefore, we discuss briefly about its key features that we believe 
could be helpful for readers, especially those who are not familiar with its
applications and internal mechanisms. 
In Fig.~\ref{fig:lock-in}, we can see a simple block diagram of a LIA, adapted from a Zurich Instruments white paper
\cite{instruments2016principles}.
The LIA allows the detection of very small signals buried inside noise.
This is possible because the multiplication stage of the LIA is performed before
the detected analog signal is digitalized.
In the conversion from analog to digital signals, the vertical resolution of the
LIA is limited by the input voltage range and by the number of bits used in the
discretization of the input analog values.
In the simplest configuration the input signal of the LIA is the response $x(t)$ of the parametric resonator to additive noise. 
This input signal is split into the two LIA channels in quadrature: a cosine and a sine channel. 
In the cosine channel the input signal is multiplied by a cosine wave at half
the pump frequency and fed to a low-pass filter (LPF),
while in the sine channel this signal is multiplied by a sine wave at half
the pump frequency and fed to a second LPF.
In the model studied here, each of the LIA LPFs consists
of only one low-pass $RC$ filter with a time constant $\tau=RC$.
The LIA outputs $x_\mathrm{L}(t)$ and $y_\mathrm{L}(t)$ are known as the cosine and sine quadratures, respectively.
The mathematical expressions for $x_\mathrm{L}(t)$ and $y_\mathrm{L}(t)$ are given by
\beq
\begin{aligned}
x_\mathrm{L}(t) &= \frac{1}{\tau} \int_{-\infty}^t \ee^{-(t - t')/\tau} \cos(\omega t') x(t')\, dt',\\
y_\mathrm{L}(t) &= \frac{1}{\tau} \int_{-\infty}^t \ee^{-(t - t')/\tau} \sin(\omega t') x(t')\, dt'.
\end{aligned}
\label{eq:x_Ly_L}
\eeq

It is important to calculate the Fourier transforms of $x_\mathrm{L}(t)$ and
$y_\mathrm{L}(t)$
given in Eq.~\eqref{eq:x_Ly_L}.
We find
\beq
\resizebox{.91\hsize}{!}{$
\begin{aligned}
\tilde x_\mathrm{L}(\nu) &= \int_{-\infty}^\infty \ee^{i\nu t}x_\mathrm{L}(t)dt=
\frac1\tau\int_{-\infty}^\infty \ee^{t'/\tau}\cos(\omega t')x(t')dt'
\int_{t'}^\infty \ee^{-(1/\tau-i\nu)t}dt
=\frac{\tilde x(\nu+\omega)+\tilde x(\nu-\omega)}{2(1-i\nu\tau)},\\
\tilde y_\mathrm{L}(\nu) &= \int_{-\infty}^\infty \ee^{i\nu t}y_\mathrm{L}(t)dt=
\frac1\tau\int_{-\infty}^\infty \ee^{t'/\tau}\sin(\omega t')x(t')dt'
\int_{t'}^\infty \ee^{-(1/\tau-i\nu)t}dt
=\frac{\tilde
x(\nu+\omega)-\tilde x(\nu-\omega)}{2i(1-i\nu\tau)}.
\end{aligned}
$}
\label{eq:x_Ly_L-nu}
\eeq
Hence, we obtain $\tilde x_\mathrm{L}(0)=\tilde x'(\omega)$ and $\tilde
y_\mathrm{L}(0)=\tilde x''(\omega)$.
This result is independent of the LIA time constant $\tau$.
\FloatBarrier
\section{The HBM analysis of degenerate parametric amplification}
\label{sec:HBM1}
We seek a stationary solution to the integro-differential equation \eqref{pa_lockin_feedback_amp} in the form
\beq
x(t)=\frac12\left[A_1(\omega)\ee^{-i\omega t}+A_1^*(\omega)\ee^{i\omega t}\right],
\eeq
where the value of the constant coefficient $A_1(\omega)$ will be determined
below.
With this substitution in Eq.~\eqref{eq:x_L} we obtain
\[
x_\mathrm{L}(t)=\frac14\left[A_1+A_1^*+\frac{A_1\ee^{-2i\omega t}}{1-2i\omega\tau}+\frac{A_1^*\ee^{2i\omega t}}{1+2i\omega\tau}\right].
\]
Using the fact that the functions $\ee^{\pm i\omega t}$ are linearly independent,
we find
\beq
\begin{aligned}
\left[1-\eta\frac{\chi(\omega)\omega\tau}{1-2i\omega\tau}\right]A_1-i\frac{\chi(\omega)}2\left(F_\mathrm{p}+\eta\right)A_1^*
&=\chi(\omega)F_\mathrm{s} \ee^{-i\varphi_0},\\
\left[1-\eta\frac{\chi^*(\omega)\omega\tau}{1+2i\omega\tau}\right]A_1^*+i\frac{\chi^*(\omega)}2\left(F_\mathrm{p}+\eta\right)A_1
&=\chi^*(\omega)F_\mathrm{s} \ee^{i\varphi_0},
\end{aligned}
\eeq
whose solution is given by
\beq
\left(
\bea{c}
A_1\\A_1^*
\ea
\right)=
\frac{F_\mathrm{s}}{\det M}
\left(
\bea{cc}
1-\eta\frac{\chi^*(\omega)\omega\tau}{1+2i\omega\tau}&i\frac{\chi(\omega)}2\left(F_\mathrm{p}+\eta\right)\\
-i\frac{\chi^*(\omega)}2\left(F_\mathrm{p}+\eta\right)&1-\eta\frac{\chi(\omega)\omega\tau}{1-2i\omega\tau}
\ea
\right)
\left(
\bea{c}
\chi(\omega)\ee^{-i\varphi_0}\\
\chi^*(\omega)\ee^{i\varphi_0}
\ea
\right),
\label{eq:A_1}
\eeq
where
\beq
\det M=\left|1-\frac{\eta\chi(\omega)\omega\tau}{1-2i\omega\tau}\right|^2-\frac{|\chi(\omega)|^2}4\left(F_\mathrm{p}+\eta\right)^2.
\label{eq:detM}
\eeq
The transition to parametric instability occurs when $\det M=0$.
In the limit $\omega\tau\rightarrow\infty$, we obtain the same expression for
the threshold in the first-order AM approximation given in
Eq.~\eqref{eq:threshold_avg}.
The expression for gain with respect to the forced harmonic oscillator
response in the HBM approximation can be written as
\beq
G_{\mathrm{HBM}}(\omega,\varphi_0)=\frac{|A_1(\omega, \varphi_0)|}{|\chi(\omega)F_\mathrm{s}|}.
\label{gain_HBM}
\eeq
From this, we find that the minimum and maximum gains are given
by
\beq
\begin{aligned}
G^{\mathrm{min}}_{\mathrm{HBM}}(\omega)&=\frac{1}{\left|1-\frac{\eta\chi(\omega)\omega\tau}{1-2i\omega\tau}\right|+\frac{|\chi(\omega)|}2\left|F_\mathrm{p}+\eta\right|},\\
G^{\mathrm{max}}_{\mathrm{HBM}}(\omega)&=\frac{1}{\left|\left|1-\frac{\eta\chi(\omega)\omega\tau}{1-2i\omega\tau}\right|-\frac{|\chi(\omega)|}2\left|F_\mathrm{p}+\eta\right|\right|}.
\end{aligned}
\label{G_minG_max}
\eeq
\section{The HBM approximation to the HB line}
In the numerical analysis of model \eqref{pa_lockin_feedback_amp2} (see Figs.
\ref{fig:threshold_FB}-\ref{fig:transient}) we found that
there is a HB to instability that was not predicted by the AM or
by the HBM.
The HBM as developed in subsection \ref{sec:HBM1} has to be modified in
order to predict this bifurcation as described below.
We seek a quasi-periodic solution to Eqs.~\eqref{pa_lockin_feedback_amp2} 
at a HB point in the following form
\beq
\begin{aligned}
x(t) &= \frac12\left[A\ee^{-\imath(\omega-\Delta)t}+A^*\ee^{\imath(\omega-\Delta)t}\right]
+\frac12\left[B\ee^{-\imath(\omega+\Delta)t}+B^*\ee^{\imath(\omega+\Delta)t}\right],\\
z(t) &= \frac12\left[C\ee^{-\imath\Delta t}+C^*\ee^{\imath\Delta t}\right],
\end{aligned}
\label{eq:quasi_periodic}
\eeq
where $\Delta$ is a detuning that will be determined below, while the
coefficients $A$, $B$, and $C$ need not be known as we are interested only in
obtaining the threshold to instability via a HB.
We find the algebraic equations
\beq
\begin{aligned}
&\left[\omega_0^2-(\omega-\Delta)^2-\imath\gamma(\omega-\Delta)\right]A-\imath\frac{F_\mathrm{p}}2B^*=\imath\eta C^*,\\
&\left[\omega_0^2-(\omega+\Delta)^2-\imath\gamma(\omega+\Delta)\right]B-\imath\frac{F_\mathrm{p}}2A^*=\imath\eta C,\\
&-\imath\Delta C=-\frac C\tau+\frac{A^*+B}{2\tau}
\implies C=\frac{A^*+B}{2(1-\imath\tau\Delta)}.
\end{aligned}
\eeq
By replacing $C$ in the above expressions, we obtain
\beq
\begin{aligned}
&\left[1 -\frac{\imath\eta \chi(\omega-\Delta)}{2(1+\imath\tau\Delta)}\right]A-
\imath\frac{\chi(\omega-\Delta)}2\left[
F_\mathrm{p}+\frac{\eta}{1+\imath\tau\Delta}\right]B^*=0
,\\
&\imath\frac{\chi^*(\omega+\Delta)}2\left[F_\mathrm{p}+\frac\eta{(1+\imath\tau\Delta)}\right]A
+\left[1+\frac{\imath\eta\chi^*(\omega+\Delta)}{2(1+\imath\tau\Delta)}\right]B^*=0.
\end{aligned}
\eeq
We find the characteristic equation (a necessary condition for nontrivial
solutions) to be
\beq
\left[1-\frac{\imath\eta \chi(\omega-\Delta)}{2(1+\imath\tau\Delta)}\right]
\left[1+\frac{\imath\eta\chi^*(\omega+\Delta)}{2(1+\imath\tau\Delta)}\right]
-\frac{\chi(\omega-\Delta)\chi^*(\omega+\Delta)}4
\left[F_\mathrm{p}+\frac{\eta}{1+\imath\tau\Delta}\right]^2
=0.
\eeq
This can be simplified to
\beq
F_\mathrm{p}^2(1+\imath\tau\Delta)+2F_\mathrm{p}\eta=4\eta\omega\left(\gamma+2\imath\Delta\right)
+\frac{4(1+\imath\tau\Delta)}{\chi(\omega-\Delta)\chi^*(\omega+\Delta)}.
\eeq
Taking the real and imaginary parts of the characteristic equation, we find
\beq
\begin{aligned}
F_\mathrm{p}^2+2\eta F_\mathrm{p}&=
4\eta\omega\gamma+\Re\left\{\frac{4}{\chi(\omega-\Delta)\chi^*(\omega+\Delta)}\right\}
-\Im\left\{\frac{4\tau\Delta}{\chi(\omega-\Delta)\chi^*(\omega+\Delta)}\right\},\\
F_\mathrm{p}^2&=\frac{8\eta\omega}\tau
+\Im\left\{\frac{4}{\tau\Delta\chi(\omega-\Delta)\chi^*(\omega+\Delta)}\right\}
+\Re\left\{\frac{4}{\chi(\omega-\Delta)\chi^*(\omega+\Delta)}\right\}.
\end{aligned}
\label{Fp_delta_system}
\eeq
We treat $F_p$ and the detuning $\Delta$ as roots to be found by solving the
above polynomial algebraic system.


%
\end{document}